\documentclass[twocolumn,iop,numberedappendix,appendixfloats]{openjournal}
\usepackage[utf8]{inputenc}

\usepackage[british]{babel}

\usepackage{amsmath}
\usepackage{amssymb}
\usepackage{graphicx}
\usepackage{bm}
\usepackage{natbib}
\usepackage[colorlinks,allcolors=blue]{hyperref}

\renewcommand{\vec}{\boldsymbol}
\newcommand{\mat}[1]{\mathrm{\mathbf{#1}}}
\newcommand{\tp}{{\mkern-0.5mu\mathsf{T}}}
\newcommand{\n}{\hat{u}}
\newcommand{\D}{\mathrm{d}}
\newcommand{\E}{\mathrm{e}}
\newcommand{\I}{\mathrm{i}}
\newcommand{\ev}[1]{\langle#1\rangle}
\DeclareMathOperator{\atanh}{atanh}

\newlength{\twocolwidth}

\begin{document}

\journalinfo{The Open Journal of Astrophysics}
\submitted{submitted XXX; accepted YYY}

\shorttitle{\emph{GLASS:} Generator for Large Scale Structure}
\shortauthors{Tessore, Loureiro, Joachimi, von Wietersheim-Kramsta \& Jeffrey}

\title{\emph{GLASS:} Generator for Large Scale Structure}

\author{Nicolas Tessore$^{\star1}$}
\author{Arthur Loureiro$^{\dagger1,2,3}$}
\author{Benjamin Joachimi$^{1}$}
\author{Maximilian von Wietersheim-Kramsta$^{1}$}
\author{Niall Jeffrey$^{1}$}

\affiliation{$^1$ Department of Physics \& Astronomy,
University College London, Gower Street, London WC1E 6BT, UK}
\affiliation{$^2$ Institute for Astronomy, University of Edinburgh, Royal
Observatory, Blackford Hill, Edinburgh EH9 3HJ, UK}
\affiliation{$^3$ Astrophysics Group and Imperial Centre for Inference and
Cosmology (ICIC), Blackett Laboratory, Imperial College London, London SW7 2AZ,
UK}
\thanks{$^\star$ E-mail: \nolinkurl{n.tessore@ucl.ac.uk} \\
$\dagger$ E-mail: \nolinkurl{arthur.loureiro@ed.ac.uk}}

\begin{abstract}
We present \emph{GLASS}, the Generator for Large Scale Structure, a new code for
the simulation of galaxy surveys for cosmology, which iteratively builds a light
cone with matter, galaxies, and weak gravitational lensing signals as a sequence
of nested shells.    This allows us to create deep and realistic simulations of
galaxy surveys at high angular resolution on standard computer hardware and with
low resource consumption.  \emph{GLASS} also introduces a new technique to
generate transformations of Gaussian random fields (including lognormal) to
essentially arbitrary precision, an iterative line-of-sight integration over
matter shells to obtain weak lensing fields, and flexible modelling of the
galaxies sector.  We demonstrate that \emph{GLASS} readily produces simulated
data sets with per cent-level accurate two-point statistics of galaxy clustering
and weak lensing, thus enabling simulation-based validation and inference that
is limited only by our current knowledge of the input matter and galaxy
properties.
\end{abstract}


\maketitle

\section{Introduction}

Simulations are an important scientific tool for current galaxy surveys.  With
increased computational and algorithmic capabilities, past and current galaxy
surveys have used simulations for complementary purposes: modelling complex
astrophysical properties \citep{2005Natur.435..629S, 2013JCAP...06..036T,
2015MNRAS.448.2987F, 2015A&C....12..109H, 2018MNRAS.473.4077P,
2019MNRAS.486.2827D, 2018MNRAS.480..800H, 2023MNRAS.519.3154H}, validating
implementations of measurement techniques and covariance matrices
\citep{2016MNRAS.456.4156K, 2016MNRAS.459.3693X, 2017ApJ...850...24T,
2018MNRAS.481.1337H, 2020ApJS..250....2V, 2022JCAP...05..002R,
2022ApJ...940...71J} and even performing inference from comparisons of data to
realistic simulation \citep{2018PhRvD..98f3511L, 2019PhRvD.100b3519T,
2019MNRAS.488.4440A, 2021MNRAS.501.4080K, 2023MLS&T...4aLT01L,
2023JCAP...02..050K}.  Thus, the ability to simulate galaxy surveys is at the
core of achieving the necessary accuracy and precision to tackle our current
challenges in contemporary cosmology.

The fundamental reason for the use of simulations in all of the above is that it
is often significantly easier to simulate a complicated model, sometimes called
forward modelling, than it is to compute its effects analytically.  For the
upcoming generation of galaxy surveys, carried out e.g.\ by \emph{Euclid}
\citep{2011arXiv1110.3193L}, \emph{Rubin} \citep{2009arXiv0912.0201L},
\emph{DESI} \citep{2019BAAS...51g..57L}, \emph{J-PAS}
\citep{2014arXiv1403.5237B}, \emph{SphereX} \citep{2014arXiv1412.4872D},
\emph{Roman} \citep{2015arXiv150303757S}, and \emph{SKA}
\citep{2020PASA...37....7S}, collectively called Stage~4 surveys, the increase
in data volume, complexity, and survey systematics will elevate the status of
simulations from important to essential.

For galaxy surveys, simulations can be broadly split into two kinds: on the one
hand, there are very large $N$-body or hydrodynamical simulations, which compute
astrophysical processes in great detail.  These simulations can, at least in
principle, model observations with as much detail as desired, and have been used
for modelling the non-linear power spectrum \citep{1996MNRAS.280L..19P,
2010MNRAS.408..300G, 2012ApJ...761..152T, 2019MNRAS.490.4826G,
2019MNRAS.488.2121C, 2021MNRAS.507.5869A} and several effects in the non-linear
power spectrum such as neutrino masses \citep{2011MNRAS.410.1647A,
2012MNRAS.420.2551B, 2016JCAP...07..053A}, intrinsic alignments of galaxies
\citep{2000MNRAS.319..649H, 2004MNRAS.347..895H, 2013MNRAS.436..819J,
    2015SSRv..193...67K, 2015MNRAS.454.2736C, 2018ApJ...853...25W,
2022PhRvD.106l3510H}, baryonic feedback \citep{2021MNRAS.502.1401M,
2021MNRAS.508.2479B, 2022MNRAS.512.3691C}, and also for providing collaborations
with a controlled data set for testing measurement techniques
\citep{2008MNRAS.391..435F, 2016MNRAS.456.4156K, 2017ApJ...850...24T,
2019arXiv190102401D}.  However, $N$-body and hydrodynamical simulations cannot
simulate everything: at the level of so-called ``subgrid physics'', they rely on
approximate descriptions of processes below the resolution of the simulations.
Overall, the computational cost of these simulations is very high, and they
usually run on dedicated infrastructure.  Although techniques such as
``cosmology rescaling'' \citep{2010MNRAS.405..143A} and ``baryon correction
models'' \citep{2015JCAP...12..049S,2019JCAP...03..020S,2020MNRAS.495.4800A}
allow changes to some cosmological parameters within a given realisation, it is
generally not the case that one can quickly compute a few thousand independent
realisations over a range of input parameters to obtain robust statistical
measures.

On the other hand, there are statistical simulations, where one generates
realisations of relevant observables directly from their (known or assumed)
statistical distributions.  These simulations can generate many realisations of
simulated surveys with great flexibility, and have been used to generate fast
and accurate galaxy mock catalogues \citep{2016MNRAS.459.3693X,
2017JCAP...10..003A, 2020MNRAS.498.2663T, 2022JCAP...05..002R} for covariance
matrix estimation \citep{2018MNRAS.476.1050B, 2018PhRvD..98b3507G,
2019ApJ...870..111Y, 2019MNRAS.485..326L, 2022A&A...665A..56L} and validation
\citep{2018PhRvD..98d3528T, 2020MNRAS.498.4060G, 2021A&A...646A.129J,
2022JCAP...08..073A, 2022MNRAS.516.5799C}, as well as simulation-based inference
\citep{2019PhRvD.100b3519T, 2021MNRAS.501..954J, 2022arXiv220611005O, 2023MLS&T...4aLT01L,
2022MNRAS.516.4111B}.  Naturally, the statistical simulations can only be as
good as the models for their distributions, and obtaining such models
theoretically is essentially the problem that we are trying to solve in the
first place.

Recently, there has been growing use of a hybrid approach to simulation,
situated between the physical and the statistical \citep{2014PDU.....3....1R,
    2017JCAP...08..035H, 2019arXiv190606630V, 2020JCAP...09..048T,
    2020PhRvD.101h2003K, 2021JOSS....6.3056A, 2022MNRAS.516.1670S,
2023ApJS..264...29A}.  Here, the idea is to make an initial statistical
simulation of some appropriate quantity that is well understood, e.g.\ the
luminosity function, and then forward-model the more difficult observables
through a series of physically inspired models.  Such models usually take some
limited input, compute some effect on said input, and produce some limited
output, which is far easier to describe than the equivalent effect on the
eventual observables.  At the same time, it reduces the necessary theoretical
modelling to the initial random sampling:  For a fixed ``amount of theory'', any
number of observations or observational effects can be taken into account simply
by combining more and more models.  This kind of simulation is therefore well
suited to likelihood-free or simulation-based inference
\citep{2018MNRAS.477.2874A, 2019MNRAS.488.4440A, 2020PNAS..11730055C, 2020arXiv201105991J,
2021MNRAS.501..954J, 2022MNRAS.511.5689H, 2023MLS&T...4aLT01L}, which is a
promising new avenue for cosmological analysis.

The idea has been applied to galaxy surveys for weak lensing by
\citet{2016MNRAS.459.3693X}.  In their approach, matter fields are generated
from a random lognormal distribution, and the weak lensing fields are
subsequently computed by a line-of-sight integration, similar to the actual
physical process of weak lensing.  Unfortunately, the exact method of
\citet{2016MNRAS.459.3693X} quickly becomes too computationally expensive.  The
matter fields are discretised as shells, in the form of \emph{HEALPix} maps
\citep{2005ApJ...622..759G} with a certain thickness in the radial direction.
For accurate numerical results, the line-of-sight intervals must be small enough
that two consecutive matter intervals remain significantly correlated.  If that
is not the case, too much of the large-scale structure is smoothed out by the
discretisation, and is subsequently missing from the weak lensing fields.  That
limits the line-of-sight intervals to be of order 100~Mpc comoving.  A
simulation up to redshift~3, which is required for many applications in Stage~4
galaxy surveys, would thus require the simultaneous generation of around 60
matter fields.  For \emph{HEALPix} maps of a given $N_{\mathrm{side}}$
parameter, this means generating $60 \times 12 \times N_{\mathrm{side}}^2$
floating point numbers.  Using $N_{\mathrm{side}} = 8192$, as necessary for
high-resolution science in Stage~4 surveys, the resulting memory requirement is
around 400~gigabyte for maps of the matter field alone.

Here, we set out to make this approach more computationally feasible for even
the largest simulations.  As stated above, our main insight is that one can
perform the entire simulation iteratively.  If only a limited number of matter
shells remain effectively correlated, as is the case for large-scale structure,
then we only need to keep that number of shells in memory.  Along the way, we
obtain many other improvements for simulating galaxy surveys, which are useful
even beyond this specific computational method.  The resulting code is modular,
extensible, and publicly available as the \texttt{glass} package for
Python.\footnote{%
Available from the Python Package Index.
}

\begin{figure}%
\centering%
\includegraphics[width=\columnwidth]{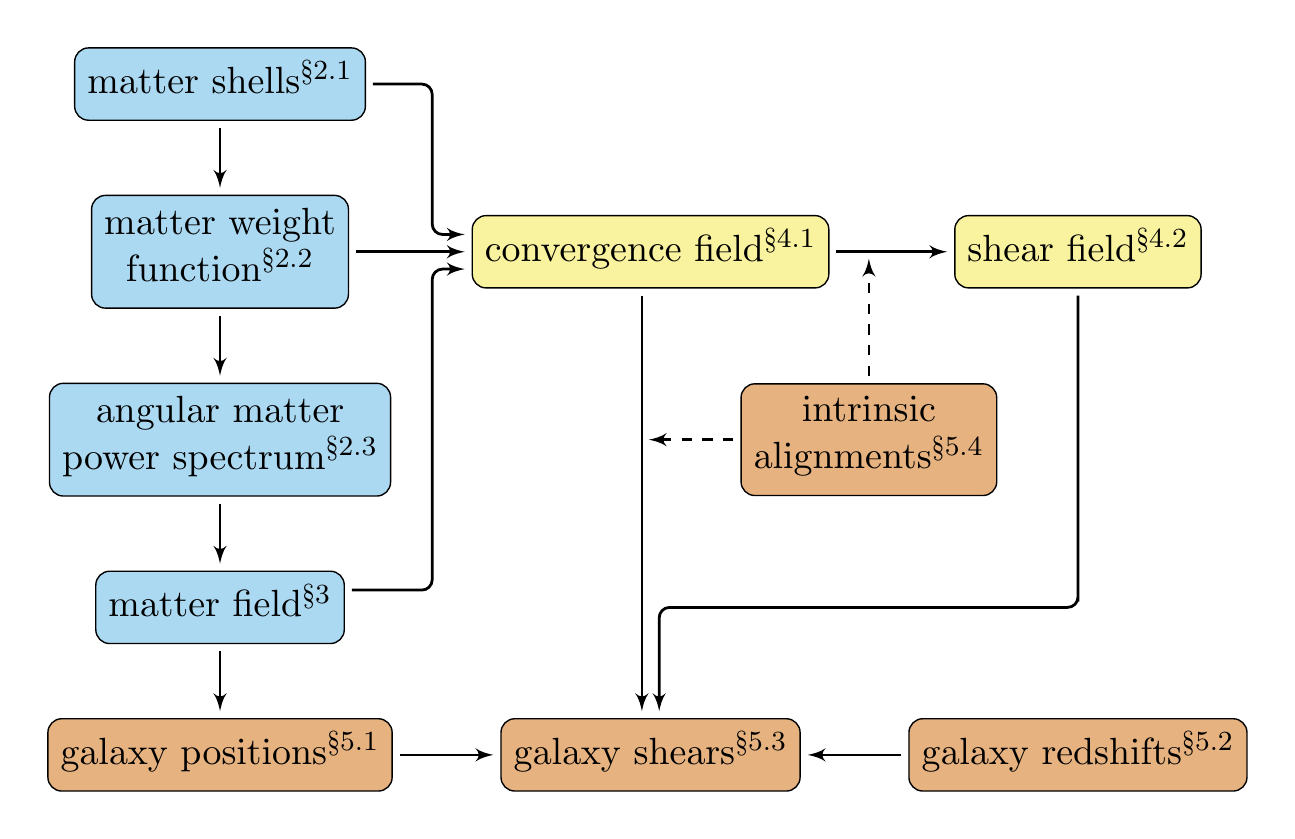}%
\caption{%
Flow chart of the typical simulation steps for a weak lensing galaxy survey.
Superscripts indicate the section where a particular step is discussed in this
work.
}%
\label{fig:flowchart}%
\end{figure}

The outline of this work mirrors the steps for simulating a weak lensing galaxy
survey, shown in Figure~\ref{fig:flowchart}.  In Section~\ref{sec:matter}, we
introduce the discretisation of the matter field into nested shells.  In
Section~\ref{sec:sampling}, we show how the matter field can be sampled
iteratively using a transformed Gaussian distribution.  In
Section~\ref{sec:lensing}, we show how the weak lensing fields, which are
integrals over all matter shells of lower redshift, can be computed iteratively
via a recurrence.  In Section~\ref{sec:galaxies}, we show how we can populate
the simulation with galaxies, as far as necessary for a cosmological galaxy
survey.  We then present an actual simulation using our models and
implementation in Section~\ref{sec:simulation}.  Finally, we discuss our results
in Section~\ref{sec:discussion}.  We provide some additional details of a more
technical nature in Appendices~\ref{sec:app-a}, \ref{sec:app-b}, and
\ref{sec:app-c}.

\section{Matter}
\label{sec:matter}

\begin{figure*}%
\centering%
\includegraphics[width=\textwidth]{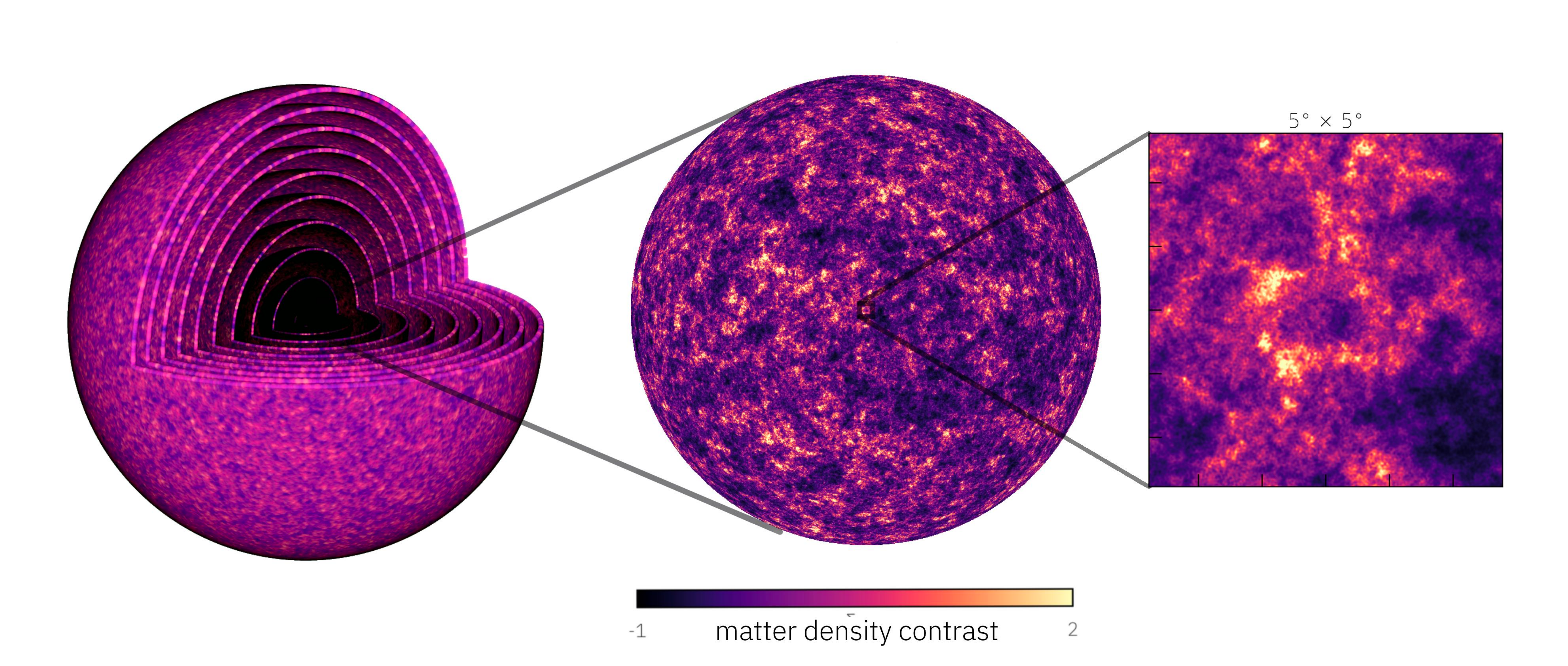}%
\caption{%
Ten shells of the discretised matter field as simulated by \emph{GLASS}, with
the first shell shown in detail. The simulations were created with
$N_{\mathrm{side}} = 8192$ ($8.05\times 10^8$ pixels) over a redshift range $0.0
\le z \le 1.0$ in 10 shells of $\Delta z = 0.1$.  Shown are an illustration of
the concentric nested matter shells at reduced resolution (\emph{left}), a
hemisphere of the innermost matter density shell at full resolution
(\emph{centre}), and a $5^{\circ} \times 5^{\circ} $ zoom into the first matter
shell showing the details in the simulated fields (\emph{right}).
}%
\label{fig:shells}%
\end{figure*}

Our overarching goal in this work is to simulate the universe as it is
accessible to a wide-field galaxy survey.  This is a universe at relatively late
times, where radiation has become insignificant, and galaxies are formed.  If
there is dark energy, it does not imprint much of an interesting signal, except
for an accelerated expansion of the cosmological background.  A galaxy survey
therefore ultimately probes matter, and particularly its spatial distribution,
the so-called large-scale structure of the universe.  But most matter appears to
be dark matter, which we cannot detect directly.  Instead, galaxy surveys
actually observe two phenomena which trace the matter distribution, and which we
must therefore ultimately simulate: weak gravitational lensing and the
distribution of galaxies.

The way we approach the simulation mirrors the real astrophysical situation.
First, we simulate the matter field itself.  We do so by means of a statistical
simulation, creating a random field with just the right spatial distribution to
look like the large-scale structure of the universe, or at least when applying
the statistics in which we are interested.  Once we have the matter field, we
then compute the associated effects of weak gravitational lensing and galaxies
using a physically inspired model.  We must hence be careful to get the matter
distribution right to a high degree of precision and accuracy, even if we do not
directly observe it, since everything else will depend on it later.  We split
the task in two:  This section treats the definition of the matter fields in our
simulation, while the next section discusses how to perform an accurate
statistical simulation.

Throughout the text, we assume a standard~$\Lambda$CDM cosmology.  We expect
that most results continue to hold in most extensions to~$\Lambda$CDM, perhaps
with some minor modification of e.g.\ the weak lensing sector.

Cosmological parameters and functions used here and in the following sections
are the matter density fraction~$\Omega_{\mathrm{m}}$, the Hubble function~$H$,
of which the present value is the Hubble constant~$H_0$, and the dimensionless
Hubble function~$E = H/H_0$.  Relevant distance functions are the comoving
distance~$d_{\mathrm{c}}$, and the transverse comoving
distance~$d_{\mathrm{M}}$.  We mainly use dimensionless distance functions in
units of the Hubble distance~$d_{\mathrm{H}} = c/H_0$, namely the dimensionless
comoving distance~$x_{\mathrm{c}} = d_{\mathrm{c}}/d_{\mathrm{H}}$, and the
dimensionless transverse comoving distance~$x_{\mathrm{M}} =
d_{\mathrm{M}}/d_{\mathrm{H}}$.  The matter distribution in the universe is
characterised by the matter density contrast~$\delta = (\rho -
\bar{\rho})/\bar{\rho}$, where $\rho$ is the matter density at a given point in
space, and~$\bar{\rho}$ is the cosmic mean matter density at that point in time.

Whenever results are computed explicitly, we must pick a specific set of
background cosmological parameters values; we use~$\Omega_{\mathrm{m}} = 0.3$
and~$H_0 = 70$~km~s$^{-1}$~Mpc$^{-1}$.

\subsection{Matter shells}

To simulate the matter distribution of the universe, we must start by picking a
suitable discretisation of three-dimensional space.  Our goal is to simulate
wide-field galaxy surveys for cosmology, and in particular those surveys that
measure weak gravitational lensing.  These surveys observe millions, and soon
billions, of individual galaxies, by taking highly resolved images of galaxy
fields.  But they do not generally observe a significant amount of galaxies by
any spectroscopic means.  It follows that the kind of galaxy survey we wish to
simulate has \emph{i)} very high angular resolution, \emph{ii)} fairly low
resolution along the line of sight.

We construct our simulation accordingly, by dividing space into a series of
nested spherical shells centred on the observer, as shown in
Figure~\ref{fig:shells}.  The shells are specified by the redshifts
\begin{equation}
    0 = z_0 < z_1 < z_2 < z_3 < \ldots
\end{equation}
of their boundaries, so that the shell with index~$i > 0$ contains redshifts~$z$
with~$z_{i-1} \le z \le z_i$.  We can thus construct shells with any desired
radial resolution.  As we will show below, using nested shells also has another
major advantage:  Any outer shell can be simulated conditional only on its inner
shells, so that we are able to iteratively construct an entire light cone, one
shell at a time.

To compute the distribution of matter over a given shell~$i$, we first fix a
radial weight function~$W_i$, which does not have to be normalised.  We then
use~$W_i$ to project the matter density contrast~$\delta$ in shell~$i$ along the
line of sight and onto the unit sphere.  This yields a spherical
function~$\delta_i$ which is the averaged matter density contrast in shell~$i$,
\begin{equation}
\label{eq:delta_i}
    \delta_i(\n)
    = \frac{\int \delta\bigl(d_{\mathrm{c}}(z) \, \n\bigr) \, W_i(z) \, \D z}
           {\int W_i(z) \, \D z} \;,
\end{equation}
where~$\n$ is a unit vector that parametrises the surface of the sphere, and the
radial direction is parametrised as usual by the redshift~$z$, so
that~$d_{\mathrm{c}}(z) \, \n$ is the three-dimensional comoving position of a
point along the line-of-sight in the direction of~$\n$.

In practice, we then need to further discretise~$\delta_i$ in the angular
dimensions, since we cannot compute with continuous functions on the sphere.  We
therefore construct a map~$\delta_{i,k} = \delta_i(\n_k)$ by evaluating the
field~$\delta_i$ over the spherical \emph{HEALPix} grid of points~$\n_k$, $k =
1, \ldots, 12 \, N_{\mathrm{side}}^2$, with~$N_{\mathrm{side}}$ a chosen
\emph{HEALPix} resolution parameter.

\subsection{Matter weight functions}
\label{sec:weight_functions}

The radial weight function~$W_i$ in the matter field~\eqref{eq:delta_i} is in
principle a free parameter of the simulation.  In this work, we assume a uniform
weight in redshift,
\begin{equation}
\label{eq:w-redshift}
    W_i(z) = \begin{cases}
        1 & \text{if $z_{i-1} \le z < z_i$,} \\
        0 & \text{otherwise.}
    \end{cases}
\end{equation}
We show in Sections~\ref{sec:lensing} and~\ref{sec:galaxies} why the uniform
weight function~\eqref{eq:w-redshift} is a good choice for simulations that
include weak gravitational lensing or galaxy distributions.

Nevertheless, there are situations in which a different choice of matter weight
function might be appropriate.  For example, instead of~\eqref{eq:w-redshift},
we could choose a uniform weight in comoving distance,
\begin{equation}
\label{eq:w-distance}
    W_i(z) = \begin{cases}
        1/E(z) & \text{if $z_{i-1} \le z < z_i$,} \\
        0 & \text{otherwise,}
    \end{cases}
\end{equation}
where~$E$ is the dimensionless Hubble function.  A true volume average of the
matter density contrast is achieved if the weight function is proportional to
the differential comoving volume,
\begin{equation}
\label{eq:w-volume}
    W_i(z) = \begin{cases}
        x_{\mathrm{M}}^2(z)/E(z) & \text{if $z_{i-1} \le z < z_i$,} \\
        0 & \text{otherwise.}
    \end{cases}
\end{equation}
Similarly, one can obtain maps of the true discretised mass by averaging the
mean matter density,
\begin{equation}
\label{eq:w-density}
    W_i(z) = \begin{cases}
        \bar{\rho}(z) \, x_{\mathrm{M}}^2(z)/E(z) &
                                            \text{if $z_{i-1} \le z < z_i$,} \\
        0 & \text{otherwise.}
    \end{cases}
\end{equation}
The weight functions~\eqref{eq:w-distance}, \eqref{eq:w-volume},
and~\eqref{eq:w-density} may therefore be good choices in simulations where
these physical quantities are of particular interest.\footnote{%
Since the matter weight function is purely a means for projecting the
three-dimensional matter distribution onto the sphere, the distribution~$n(z)$
of eventually observed sources is generally not a good choice.
}

\subsection{Angular matter power spectra}

In principle, the discretised matter fields~\eqref{eq:delta_i} can be provided
from any suitable source.  For example, it is possible to compute the matter
density contrast~$\delta_i$ in each shell from the outputs of an $N$-body
simulation.  Of course, we will normally want to generate the matter field as
part of our simulation, and it must therefore contain the information that is
relevant for cosmology.  For the wide-field galaxy surveys we wish to simulate,
that means we have to imprint the correct two-point statistics.

The two-point statistics of our generated matter fields are described by the
angular matter power spectrum for each pair of shells.  Many of the usual
cosmology codes such as \emph{CAMB} \citep{2000ApJ...538..473L,
2002PhRvD..66j3511L}, \emph{CCL} \citep{2019ApJS..242....2C}, or \emph{CLASS}
\citep{2011arXiv1104.2932L, 2011JCAP...07..034B} can compute these spectra,
which only requires the matter weight function~${W}_i$ that defines the matter
field~\eqref{eq:delta_i} in each shell~$i$.  Since~$\delta_i$ is the projection
of the matter field, and not the galaxy field, the angular matter power spectrum
is computed without bias, redshift-space distortions, or any other such
observational effect.

This is important, because the angular power spectra completely determine the
underlying physical model for matter in the simulation.  If the angular power
spectra are computed e.g.\ using only the linear matter power spectrum, the
simulation will only produce the linear matter field.  Similarly, if the
angular power spectra include a full non-linear treatment of matter, so will
the simulation.  The only task of the simulation is to reproduce the given
angular power spectra faithfully, which we achieve using the methods of the
next section.

The fact that we consider many relatively thin shells with a thickness
of~$\Delta z \lesssim 0.1$ in redshift means that the computation of the angular
power spectra must largely be performed without use of Limber's approximation
\citep{1953ApJ...117..134L, 1998ApJ...498...26K, 2007A&A...473..711S}.  For this
work, we use \emph{CAMB}, since it is widely available, and allows Limber's
approximation to be switched off altogether.  To work around a numerical issue
in \emph{CAMB} for flat matter weight functions that do not go to zero at $z =
0$, we slightly modify~\eqref{eq:w-redshift} to increase linearly from zero at
$z = 0$ to unity at $z = 0.1$, which is an otherwise negligible change.  To
obtain results at the required level of accuracy, we also set the
\texttt{TimeStepBoost} parameter in \emph{CAMB} to~5.

\section{Sampling random fields on the sphere}
\label{sec:sampling}

The projected matter field of the previous section is at the heart of our
simulations, as we will derive the weak gravitational lensing fields and the
distribution of galaxies from the matter shells in the following sections.  In
this section, we show how we can produce random realisations of the projected
matter density contrast~\eqref{eq:delta_i} with
\begin{itemize}
    \item[i)] a realistic distributions of values of the matter field, i.e.\ the
        one-point statistics, and
    \item[ii)] the physically correct angular matter power spectrum, i.e.\ the
        two-point statistics.
\end{itemize}
These two criteria are imposed by our aim of producing simulations for the
typical clustering and weak lensing studies done on wide-field galaxy surveys.

Sampling a Gaussian random map~$X$ with fully specified statistical properties
is readily done.  However, the normal distribution is not a good model for the
evolved matter fields that we wish to simulate.  But if we apply a suitable
transformation~$f$ to the map, we obtain a second random map~$Y = f(X)$ which
now has a different distribution.  By picking the right transformation, we will
be able to recreate the one-point statistics of the matter field with high
fidelity.  The main challenge is then to imprint the correct two-point
statistics onto the transformed map~$Y$ via the transformation~$f(X)$.

There is also a computational reason for basing our simulation on Gaussian
random maps.  The random realisations must contain the right correlations
between the projected matter fields across all simulated shells.  This means
that we must either simulate, and hence hold in memory, all shells at once, or
we must sample each new shell conditional on the existing shells.  The former is
usually not feasible for high-resolution maps without dedicated hardware.  But
the latter is particularly simple for Gaussian random maps.

\subsection{Transformed Gaussian random fields}

Let us for the moment assume that the transformation~$f$ has already been fixed.
Naturally, we must match the distribution of the Gaussian map~$X$ to the desired
distribution of the transformed map~$Y$, such that the realisation has e.g.\ the
correct mean and variance after the transformation.  In the following, we always
assume that the fields are homogeneous, i.e.\ invariant under rotations, as
asserted by the cosmological principle.  If the Gaussian map~$X$ is homogeneous,
it has the same mean~$\mu$ and variance~$\sigma^2$ everywhere, in the sense that
for all points~$\n$ on the sphere the expectation over realisations, denoted
by~$\ev{\,\cdot\,}$, is
\begin{equation}
    \ev{X(\n)} = \mu \quad \text{and} \quad
    \ev{X^2(\n)} - \mu^2 = \sigma^2 \;.
\end{equation}
Since~$Y(\n) = f\bigl(X(\n)\bigr)$ and~$X(\n)$ is normally distributed with
mean~$\mu$ and variance~$\sigma^2$, it follows that the transformation~$Y =
f(X)$ of a homogeneous Gaussian map~$X$ remains homogeneous, and the
distribution of~$Y$, and thus all one-point statistics, depend solely on~$f$,
$\mu$, and~$\sigma^2$.  In particular, $Y$ has the same mean~$\ev{Y}$ and
variance~$\ev{Y^2} - \ev{Y}^2$ everywhere.

Apart from the overall distribution of the values, the transformation must also
imprint the realised map~$Y$ with the correct two-point statistics, since that
is where we extract cosmological information from the simulations.  If~$Y$
and~$Y'$ are two not necessarily distinct homogeneous spherical random fields,
the correlation in the respective points~$\n$ and~$\n'$ is described by the
angular correlation function~$C$,
\begin{equation}
\label{eq:cov-y}
    \ev{Y(\n) \, Y'(\n')}
    = C(\theta) \;,
\end{equation}
which, due to homogeneity, is a function of the angle~$\theta$ between~$\n$
and~$\n'$ alone.  Let both fields be the respective transformations~$Y = f(X)$
and~$Y' = f'(X')$ of homogeneous Gaussian fields~$X$ and~$X'$, so that~$X(\n)$
and~$X'(\n')$ are jointly normal with the respective means~$\mu$ and~$\mu'$ and
variances~$\sigma^2$ and~$\sigma^{\prime2}$.  If the correlations between~$X$
and~$X'$ are given by the correlation function~$G$,
\begin{equation}
    \ev{X(\n) \, X'(\n')}
    = G(\theta) \;,
\end{equation}
the joint distribution of~$X(\n)$ and~$X'(\n')$, being jointly normal random
variables, is completely described by the values of~$\mu$, $\mu'$, $\sigma^2$,
$\sigma^{\prime2}$, and~$G(\theta)$.  It follows that the
correlation~\eqref{eq:cov-y} between the transformed random variables~$Y(\n) =
f(X(\n))$ and~$Y'(\n') = f'(X(\n'))$ must be a function of these variables
alone,
\begin{equation}
\label{eq:gttoct}
    C(\theta)
    = C\bigl(G(\theta); \mu, \mu', \sigma^2, \sigma^{\prime2}\bigr) \;,
\end{equation}
where the form of this function depends on the transformations~$f$ and~$f'$
between the fields.  The function~$C$ will normally be obtained by
computing~\eqref{eq:cov-y} explicitly.  Inverting the result, either
analytically or numerically, then yields the function
\begin{equation}
\label{eq:cttogt}
    G(\theta)
    = G\bigl(C(\theta); \mu, \mu', \sigma^2, \sigma^{\prime2}\bigr) \;,
\end{equation}
which characterises the two-point statistics of the Gaussian maps in terms of
the two-point statistics of their transformations.

Given a transformation~$f$, we can hence expect to also be given the
relations~\eqref{eq:gttoct} and~\eqref{eq:cttogt} for translating the desired
correlations~$C(\theta)$ of~$Y$ into the correlations~$G(\theta)$ to be
imprinted onto the Gaussian random field~$X$.

\subsection{Lognormal fields}

One popular choice of transformation~$f$ for matter fields is the lognormal
distribution \citep[e.g.][]{1991MNRAS.248....1C, 2001ApJ...561...22K,
2011A&A...536A..85H, 2016MNRAS.459.3693X},
\begin{equation}
\label{eq:lognorm}
    f(x)
    = \lambda \, (\E^x - 1) \;,
\end{equation}
where the parameter~$\lambda$ is the so-called ``shift'' of the lognormal
distribution.  Since the exponential is limited to positive values, the value
of~$\lambda$ is effectively the lower bound of variates of the distribution.  A
volume devoid of any matter has matter density contrast~$\delta = -1$, so a
shift parameter~$\lambda = 1$ is usually assumed for matter fields.

The simulation of lognormal random fields on the sphere was discussed in detail
by \citet{2016MNRAS.459.3693X}, and we only repeat the
relations~\eqref{eq:gttoct} and~\eqref{eq:cttogt} here,
\begin{align}
\label{eq:lognorm-gttoct}
    C(\theta)
    &= \alpha \alpha' \Bigl\{\E^{G(\theta)} - 1\Bigr\} \;,
\\
\label{eq:lognorm-cttogt}
    G(\theta)
    &= \ln\Bigl\{1 + \frac{C(\theta)}{\alpha \alpha'}\Bigr\} \;,
\end{align}
which are characterised by the parameter~$\alpha = \ev{Y} + \lambda$ for~$Y$,
and similarly~$\alpha'$ for~$Y'$.

Lognormal distributions are widely used not only for simulating the matter field
\citep{1991MNRAS.248....1C, 2017PhRvD..96l3510B, 2016MNRAS.455.3871A,
2022JCAP...08..073A} but also weak lensing convergence fields
\citep{2011A&A...536A..85H, 2017MNRAS.466.1444C, 2017MNRAS.470.3574G,
2020MNRAS.498.4060G}.  In particular, \cite{2022PhRvD.105l3527H} showed that
lognormal distributions reproduce, up to reasonable precision and accuracy, the
bispectrum (i.e.\ three-point statistics) and the covariance (i.e.\ four-point
statistics) of the underlying fields when compared to results obtained from
$N$-body simulations over the typical scales for a Stage 4 photometric galaxy
survey.  However, the agreement between lognormal and $N$-body simulations for
higher-order statistics is not perfect, and it is conditional on the scales and
configurations analysed \citep{2023MNRAS.520..668P}.

\subsection{Gaussian angular power spectra}
\label{sec:gaussian-cl}

Having obtained a suitable transformation~$f$, such as e.g.\ the lognormal
transformation~\eqref{eq:lognorm}, and derived its relations~\eqref{eq:gttoct}
and~\eqref{eq:cttogt} for the two-point statistics, we face two further issues
before we can actually sample the Gaussian random map~$X$:  Firstly, theoretical
calculations usually do not produce~$C(\theta)$, but instead the angular matter
power spectrum~$C_l$ for the matter fields~\eqref{eq:delta_i}.  And secondly,
the procedure for sampling a Gaussian random map also requires the Gaussian
angular power spectrum~$G_l$ instead of~$G(\theta)$.  We must therefore convert
between the angular correlation functions and angular power spectra.

The conversion is done using the well-known transforms between angular
correlation functions and angular power spectra,
\begin{equation}
\label{eq:cltoct}
    C(\theta)
    = \sum_{l=0}^{\infty} \frac{2l+1}{4\pi} \, C_l \, P_l(\cos\theta) \;,
\end{equation}
with~$P_l$ the Legendre polynomial of degree~$l$, and
\begin{equation}
\label{eq:cttocl}
    C_l
    = 2\pi \int_{0}^{\pi} \! C(\theta) \,
                                P_l(\cos\theta) \sin(\theta) \, \D\theta \;,
\end{equation}
and similarly for~$G(\theta)$ and~$G_l$.  In theory, the steps to obtain~$G_l$
from~$C_l$ are hence straightforward:
\begin{itemize}
    \item[i)] Compute the correlations~$C(\theta)$ from~$C_l$
        using~\eqref{eq:cltoct},
    \item[ii)] apply relation~\eqref{eq:cttogt} to obtain~$G(\theta)$
        from~$C(\theta)$, and
    \item[iii)] compute~$G_l$ from from~$G(\theta)$ using~\eqref{eq:cttocl}.
\end{itemize}
Overall, the computation can be summarised as
\begin{equation}
\label{eq:cltogl}
    C_l \to C(\theta) \to G(\theta) \to G_l \;,
\end{equation}
which we call the ``backward'' sequence.  This name is owed to the fact that the
sampling of a Gaussian random field from~$G_l$ and subsequent transformation~$Y
= f(X)$ instead correspond to
\begin{equation}
\label{eq:gltocl}
    G_l \to G(\theta) \to C(\theta) \to C_l \;,
\end{equation}
which we consequently call the ``forward'' sequence.

\begin{figure}%
\centering%
\includegraphics[scale=0.85]{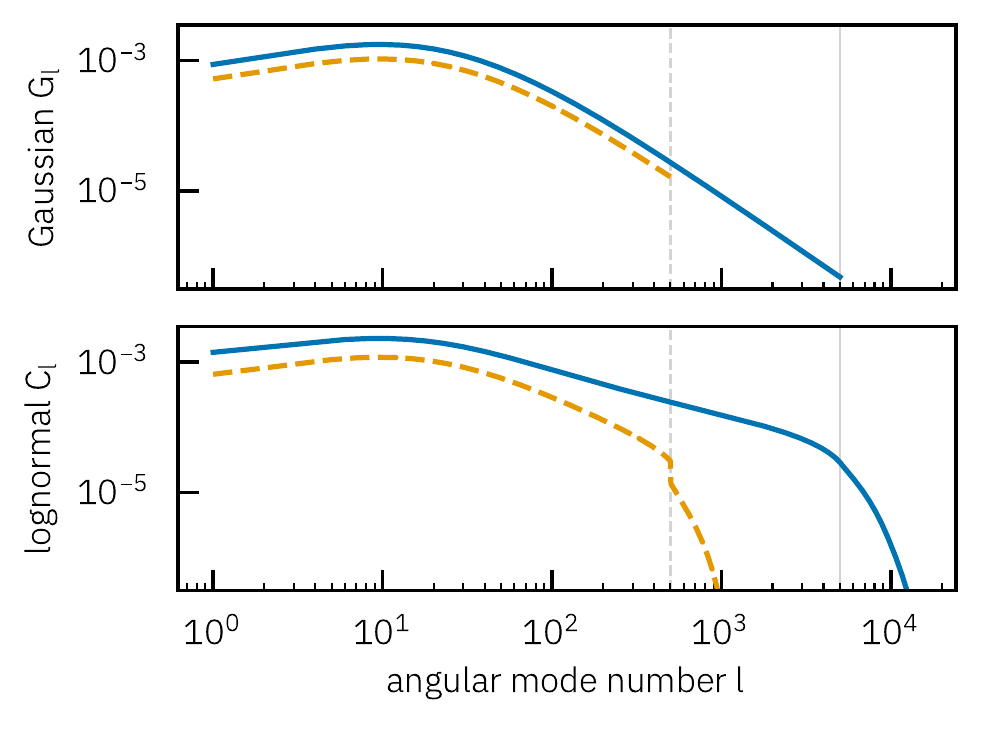}%
\caption{%
The effect of a band limit in the Gaussian angular power spectrum~$G_l$ on the
transformed angular power spectrum~$C_l$, here for the lognormal transformation
of two inputs (\emph{solid}, \emph{dashed}) with different band limits
(\emph{vertical lines}).  The shape of~$C_l$ depends critically on the band
limit of~$G_l$, and will generally have a higher band limit.
}%
\label{fig:gaussian_cl}%
\end{figure}

\begin{figure}%
\centering%
\includegraphics[scale=0.85]{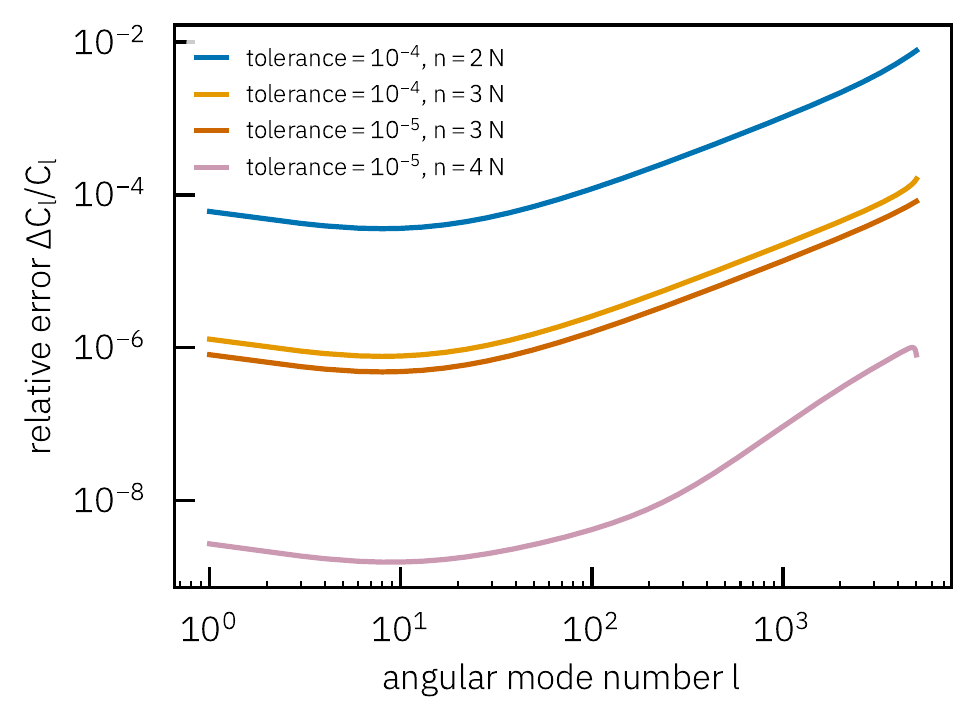}%
\caption{%
Relative error~$\Delta C_l/C_l$ of the realised angular power spectra using the
Gauss--Newton solver.  Shown are various settings of the nominal tolerance of
the algorithm and the length~$n$ of the internal Legendre transforms relative to
the length~$N$ of the inputs.
}%
\label{fig:gaussian_cl_err}%
\end{figure}

In practice, we can usually neither evaluate the infinite sum
in~\eqref{eq:cltoct} for all~$\theta$, nor the continuous integral
in~\eqref{eq:cttocl} for all~$l$, and we always have to work with angular power
spectra of finite length.  But imposing a band limit on both~$C_l$ and~$G_l$ is
problematic:  \citet{2016MNRAS.459.3693X} noted that, for lognormal fields, a
band-limited~$G_l$ yields values~$C_l$ beyond the band limit, and the same holds
more generally for any non-linear transformation~$f$.  The effect is shown in
Figure~\ref{fig:gaussian_cl}.

To work around the finite nature of their transforms, the approach of
\citet{2016MNRAS.459.3693X} was to take a given band-limited~$C_l$ and
compute~$G_l$ using the backward sequence~\eqref{eq:cltogl} at a higher band
limit.  This is shown to achieve per-cent level fidelity of the realisation when
the band limit is set very generously, which is computationally expensive, since
the cost of a discrete spherical harmonic transform increases with the square of
the band limit.  It also requires regularisation of the transformed angular
power spectra, which may at least partly be due to the fact that~$C_l$ contains
zeros when padded to a higher band limit, rendering the conversion between~$G_l$
and~$C_l$ ill-defined.

On closer inspection, the difficulty arises from use of the backward
sequence~\eqref{eq:cltogl} for directly computing~$G_l$ from a given
band-limited~$C_l$.  But there are other ways to approach the conversion
\citep{SHIELDS2011511}.  For example, we can try and solve the inverse problem
instead, which is: find a band-limited Gaussian angular power spectrum~$G_l$ of
length~$N$ such that the forward sequence~\eqref{eq:gltocl} recovers~$N$ given
values~$C_l$.  As it turns out, that approach is both simpler and more accurate.
All it needs is a standard numerical method for the solution, or approximate
solution, of non-linear equations.  Here, we use the Gauss--Newton algorithm.

To start, let~$G_l$ be an initial guess for the Gaussian angular power spectrum,
and let~$F_l$ be the residuals of the forward sequence~\eqref{eq:gltocl} and
given values~$C_l$.  The Gauss--Newton update moves from~$G_l$ to~$G_l + S_l$,
where the step~$S_l$ is found by solving the matrix equation
\begin{equation}
\label{eq:df_dg_x}
    \sum_{l'} \frac{\partial F_l}{\partial G_{l'}} \, S_{l'}
    = -F_l \;.
\end{equation}
Applying the derivative to the forward sequence~\eqref{eq:gltocl} yields
\begin{equation}
\label{eq:df_dg}
    \frac{\partial F_l}{\partial G_{l'}}
    = 2\pi \int_{0}^{\pi} \! \frac{\partial C(\theta)}{\partial G(\theta)} \,
        \frac{\partial G(\theta)}{\partial G_{l'}} \,
        P_l(\cos\theta) \sin(\theta) \, \D\theta \;.
\end{equation}
Note that~$\partial C(\theta)/\partial G(\theta)$ is the derivative
of~\eqref{eq:gttoct} with respect to~$G(\theta)$; for short, let~$\partial
C(\theta)/\partial G(\theta) = \dot{C}(\theta)$.  Like~$C(\theta)$ itself, the
function~$\dot{C}$ is characteristic of the transformation~$f$, and can be
computed.  The other derivative in~\eqref{eq:df_dg} is readily found
using~\eqref{eq:cltoct},
\begin{equation}
\label{eq:dg_dg}
    \frac{\partial G(\theta)}{\partial G_{l'}}
    = \frac{2l'+1}{4\pi} \, P_{l'}(\cos\theta) \;.
\end{equation}
Using~\eqref{eq:df_dg} and~\eqref{eq:dg_dg}, the matrix
equation~\eqref{eq:df_dg_x} becomes the integral
\begin{equation}
\label{eq:dc_s_f}
    2\pi \int_{0}^{\pi} \! \dot{C}(\theta) \, S(\theta) \,
                                P_l(\cos\theta) \sin(\theta) \, \D\theta
    = -F_l \;,
\end{equation}
where we have exchanged summation and integration to transform~$S_l$
into~$S(\theta)$ using~\eqref{eq:cltoct},
\begin{equation}
    \sum_{l'} \frac{2l'+1}{4\pi} \, S_{l'} \, P_{l'}(\cos\theta)
    = S(\theta) \;.
\end{equation}
Since the resulting equation~\eqref{eq:dc_s_f} itself is precisely the
transform~\eqref{eq:cttocl}, we obtain the result that the Gauss--Newton step
must obey $\dot{C}(\theta) \, S(\theta) = -F(\theta)$ in real space.  The
solution of~\eqref{eq:df_dg_x} therefore has the representation
\begin{equation}
\label{eq:gn-sol}
    S(\theta)
    = -\frac{F(\theta)}{\dot{C}(\theta)} \;,
\end{equation}
which can be transformed back into~$S_l$ using~\eqref{eq:cttocl}.  It only
remains to find an initial guess for the values~$G_l$, which we do using the
backward sequence~\eqref{eq:cltogl} for the fixed length~$N$.  This generally
yields a starting point such that the Gauss--Newton algorithm converges in just
a handful of iterations.

Solving for a Gaussian angular power spectrum~$G_l$ with the above method still
involves the transforms~\eqref{eq:cltoct} and~\eqref{eq:cttocl}, so that the
true, continuous transforms must in practice still be approximated by finite,
discrete ones.  The crucial difference is that we do not transform~$C$ and~$G$
here, but instead~$F$ and~$S$.  Depending on the desired accuracy, we can choose
an arbitrarily large length~$n \gg N$ for the transforms;  since they are
internal to the Gauss--Newton step, both~$C_l$ and~$G_l$ remain of the
length~$N$ that we ultimately want to realise.  As mentioned earlier, this
quadratically improves the sampling performance over methods relying on padded
spectra.

In practical terms, we note that the transforms~\eqref{eq:cltoct}
and~\eqref{eq:cttocl} are effectively discrete Legendre expansions with slightly
modified coefficients.  We can compute them using the method we describe in
Appendix~\ref{sec:app-a}, which maps~$n$ values~$F_l$ to~$n$ values~$F(\theta)$
over a regular grid of~$\theta$ values using the Fast Fourier Transform.  The
mapping is one-to-one and invertible, so that we can transform back and forth
without loss of information.  Commonly used methods based on Gaussian
quadrature, as well as the method of \citet{1994AAM...15...202D}, or the method
of \citet{2003JFAA....9..341H} used by \citet{2016MNRAS.459.3693X}, map
between~$n$ values of~$F_l$ and~$2n$ values of~$F(\theta)$, and are therefore
clearly not generally invertible.  Our transforms are very fast and do not
construct any large matrices, so that values of e.g.\ $n > 100\,000$ are readily
achievable.

To give an idea of the accuracy of our new method for computing Gaussian angular
power spectra, Figure~\ref{fig:gaussian_cl_err} shows the relative error of the
lognormal transformation of a typical angular power spectrum~$C_l$ with~$l \le
5\,000$, i.e.\ $N = 5\,001$.  We show the solution of the Gaussian angular power
spectrum~$G_l$ using a number of settings for the nominal tolerance of the
Gauss--Newton algorithm, as well as different lengths~$n$ of the internal
Legendre transforms.  To compare the result to the input, we compute the forward
sequence~\eqref{eq:gltocl} for each solution~$G_l$ using~$1\,000\,000$ terms in
the Legendre expansion.  We find that, in the regime shown, the accuracy of the
solution depends mainly on~$n$.  We adopt a tolerance of~$10^{-5}$ and $n = 3N$
as good default values, having a relative error better than~$10^{-4}$
everywhere, with the understanding that better accuracy is readily available.

Overall, this new method allows us to simulate transformed Gaussian random
fields on the sphere in such a way that the first~$N$ modes of the angular power
spectrum match any given values~$C_l$ reliably. In principle, we could therefore
accurately simulate maps of the matter fields up to the band limit~$l_{\max}$ of
a \emph{HEALPix} map, which for a given resolution parameter~$N_{\mathrm{side}}$
is $l_{\max} = 3 N_{\mathrm{side}} - 1$.  However, as shown in
Figure~\ref{fig:gaussian_cl}, the transformed random field will in general not
be band-limited to~$l_{\max}$.  Even if the angular power spectrum is simulated
accurately up to~$l_{\max}$, it is hence difficult to actually use this part of
the spectrum for practical purposes, due to aliasing from modes beyond the band
limit.  To obtain interpretable results, we find values of~$N$ somewhere
between~$N_{\mathrm{side}}$ and~$2 N_{\mathrm{side}}$ most reliable.

Because they are generally useful beyond this specific work, we provide our
implementations of the transforms~\eqref{eq:cltoct} and~\eqref{eq:cttocl} as the
stand-alone \texttt{transformcl} package for Python, and our solver for Gaussian
angular power spectra as the stand-alone \texttt{gaussiancl} package for
Python.\footnote{%
Both available from the Python Package Index.
}

\subsection{Zero monopoles}

Computer codes often produce theoretical angular matter power spectra with a
vanishing monopole.  For the simulated matter shells, this is problematic for
two reasons: Physically, it is not the case that a matter shell of finite size
has an exactly vanishing average density contrast with no variance at all.  And
mathematically, a vanishing monopole results in an ill-defined Gaussian
transformation.  The first issue requires better theoretical computations, which
is not part of our work.  But we can try and mitigate the second issue
ourselves.

More specifically, the problem is that a vanishing monopole value~$C_0 = 0$ in
the transformed angular power spectrum will generally result in a negative
monopole~$G_0$ in the Gaussian angular power spectrum.  This occurs because the
transformation mixes Gaussian modes with non-zero random values from beyond the
monopole into the monopole of the transformed field.  To counteract the
randomness, at least formally, a negative variance is required, and the Gaussian
random field becomes ill-defined.

To work around this issue, we can exclude both monopoles~$C_0$ and~$G_0$ from
our solver, fixing~$G_0 = 0$.  After the transformation, the realised field will
have a value~$C_0 > 0$ that is realistic, but arbitrary.  The result is
essentially a smooth extrapolation to~$l = 0$ of the given modes~$C_l$ with~$l >
0$, which is the best we can do to obtain a well-defined random field.

The Gauss--Newton solver is readily adapted to ignore~$C_0$ and fix~$G_0$ to its
initial value:  The latter is equivalent to~$S_0$ in the update
step~\eqref{eq:gn-sol} being zero, and there always exists a value of~$C_0$ such
that this is the case.  Since the given~$C_0$ is ignored, we can arbitrarily
assume that it was that particular value.  To obtain the constrained solution,
it therefore suffices to set~$S_0 = 0$ and~$F_0 = 0$ in the unconstrained
solution.

\subsection{Sampling the Gaussian random fields}

To sample a Gaussian random field~$X$ on the sphere with a given angular power
spectrum~$G_l$, we sample the complex-valued modes~$a_{lm}$ of its spherical
harmonic expansion,
\begin{equation}
\label{eq:alm-def}
    X(\n)
    = \sum_{lm} a_{lm} \, Y_{lm}(\n) \;.
\end{equation}
We can obtain a number of conditions on the distribution of the~$a_{lm}$.  If
the field is homogeneous, i.e.\ invariant under rotations, the mean of the modes
with~$l > 0$ must vanish,
\begin{equation}
\label{eq:alm-mean}
    \ev{a_{lm}} = 0 \;.
\end{equation}
If the field also has zero expectation, as is the case for the matter density
contrast, the same holds for the monopole~$l = 0$.  The angular power spectrum
determines the covariance of the modes with numbers~$l, m$ and~$l', m'$,
\begin{equation}
\label{eq:alm-cov}
    \ev{a_{lm} \, a_{l'm'}^*}
    = \delta^{\mathrm{K}}_{ll'} \, \delta^{\mathrm{K}}_{mm'} \, G_l \;,
\end{equation}
where the Kronecker delta expresses that differently-numbered modes are
uncorrelated, which follows from homogeneity of the field.  For a real-valued
field, the symmetry~$a_{lm}^* = (-1)^m \, a_{l,-m}$ and the
covariance~\eqref{eq:alm-cov} together imply that the pseudo-variance of the
modes vanishes for~$m \ne 0$,
\begin{equation}
\label{eq:alm-pvar}
    \ev{a_{lm}^2}
    = (-1)^m \, \ev{a_{lm} \, a_{l-m}^*}
    = \delta^{\mathrm{K}}_{m0} \, G_l \;.
\end{equation}
Finally, since any linear combination of normal random variables remains
normally distributed, we can sample the modes~$a_{lm}$ themselves as complex
normal random variables.

The sampling is most easily done by splitting each~$a_{lm}$ into its real and
imaginary part,
\begin{equation}
    a_{lm} = x_{lm} + \I \, y_{lm} \;,
\end{equation}
and sampling the set of~$x_{lm}$ and~$y_{lm}$ as a real-valued multivariate
normal random variable.  If the field is real-valued, the symmetry~$a_{lm}^*
=(-1)^m \, a_{l,-m}$ implies that only the~$x_{lm}$ and~$y_{lm}$ with~$m \ge 0$
need to be sampled.  By condition~\eqref{eq:alm-mean}, the means of all~$x_{lm}$
and~$y_{lm}$ vanish,
\begin{equation}
    \ev{x_{lm}} = \ev{y_{lm}} = 0 \;.
\end{equation}
By conditions~\eqref{eq:alm-cov} and~\eqref{eq:alm-pvar}, a pair of~$x_{lm}$
and~$y_{lm}$ with~$m > 0$ is uncorrelated, $\ev{x_{lm} \, y_{lm}} = 0$, with
equal variance,
\begin{equation}
    \ev{x_{lm}^2} = \ev{y_{lm}^2} = \frac{G_l}{2}
    \qquad (m > 0) \;.
\end{equation}
For~$m = 0$, the same conditions imply that
\begin{equation}
    \ev{x_{l0}^2} = G_l  \quad \text{and} \quad \ev{y_{l0}^2} = 0 \;,
\end{equation}
and thus~$y_{l0} = 0$ identically.  Furthermore, by
condition~\eqref{eq:alm-cov}, the~$x_{lm}$ and~$y_{lm}$ are pairwise
uncorrelated for different modes.  We therefore only have to sample for~$m \ge
0$ each pair of~$x_{lm}$ and~$y_{lm}$ independently, with zero mean and the
correct variance.  After an inverse spherical harmonic transform, we obtain the
Gaussian random field~$X$ with the prescribed statistics.

When correlated Gaussian random fields~$X^i$ and~$X^j$ are simulated, with~$i$
and~$j$ some indices, there is an additional condition that the covariance of
their respective modes~$a_{lm}^i$ and~$a_{lm}^j$ recovers the angular
cross-power spectrum~$G_l^{ij}$,
\begin{equation}
    \ev{a_{lm}^i \, a_{l'm'}^{j*}}
    = \delta^{\mathrm{K}}_{ll'} \, \delta^{\mathrm{K}}_{mm'} \, G_l^{ij} \;.
\end{equation}
For fixed values of~$l$ and~$m$, the sets~$\vec{x}_{lm} = \{x_{lm}^1, x_{lm}^2,
\ldots\}$ and~$\vec{y}_{lm} = \{y_{lm}^1, y_{lm}^2, \ldots\}$ taken over
different fields are thus multivariate normal random vectors with covariance
matrix
\begin{gather}
\label{eq:xlm-ylm-cov}
    \ev{x_{lm}^i \, x_{lm}^j}
    = \ev{y_{lm}^i \, y_{lm}^j}
    = \frac{G_l^{ij}}{2} \;,
    \quad m > 0 \;, \\
    \ev{x_{l0}^i \, x_{l0}^j} = G_l^{ij} \quad \text{and} \quad
    \ev{y_{l0}^i \, y_{l0}^j} = 0 \;,
\end{gather}
and remain independent across different modes.  For~$n$ correlated Gaussian
random fields, we thus have to sample the multivariate normal random
variables~$\vec{x}_{lm}$ and~$\vec{y}_{lm}$ for each~$l, m$ independently from
their $n \times n$ covariance matrix.

For our specific application, this is problematic.  At the highest map
resolutions, it is not feasible to sample the integrated matter fields for
hundreds of shells all at once, due to the amount of memory required.  However,
it is possible to sample multivariate normal random variables iteratively, which
in our case means: shell by shell.  The technique, shown in
Appendix~\ref{sec:app-b}, allows us to generate each new integrated matter field
in turn, while still imprinting the correct correlations with previous shells.
In addition, we use that the correlations of the matter field along the line of
sight become negligible above a certain correlation length, of the order of
100~Mpc.  As we show in the appendix, the iterative sampling then only requires
us to store those fields which are effectively still correlated, so that we are
able to sample arbitrarily many shells without increasing our memory
requirements.  Only the thickness of the shells determines the amount of
correlation between them, and thus how many previous shells we must store.  We
show how an informed choice can be made in Section~\ref{sec:simulation}.

\section{Weak gravitational lensing}
\label{sec:lensing}

We now use our realisation of the matter fields in each shell to compute other,
related fields, namely the convergence and shear of weak gravitational lensing.
The fact that we compute lensing from matter in deterministic fashion, close to
the real physical situation, means that we do not have to make any additional
assumptions about e.g.\ the statistical distributions of the fields.

On the other hand, it also means we have to overcome two associated
difficulties:  First and foremost, the fact that we wish to continue sampling
the fields iteratively, shell by shell.  Lensing happens continuously between
source and observer, and the computation of the lensing fields requires an
integral over the line of sight.  We therefore have to develop a way to perform
the computation iteratively.  The second difficulty is also related to the
integration:  the matter fields that we sample are already discretised into
shells, and we have to approximate the lensing integral using the existing
discretisation.

\subsection{Convergence}
\label{sec:convergence}

We compute the convergence field~$\kappa$ from the matter density
contrast~$\delta$ in the Born approximation, i.e.\@ along an undeflected line of
sight.  In the case of weak lensing, this approximation is sufficient even for
upcoming weak lensing surveys \citep{2017PhRvD..95l3503P}.  The convergence for
a source located at angular position~$\n$ and redshift~$z$ is hence
\citep[see e.g.][]{2006glsw.conf.....M}
\begin{multline}
\label{eq:kappa}
    \kappa(\n; z) \\
    = \tfrac{3\Omega_m}{2}
        \int_{0}^{z} \! \delta\bigl(d_{\mathrm{c}}(z') \, \n\bigr) \,
        \tfrac{x_{\mathrm{M}}(z') \, x_{\mathrm{M}}(z', z)}{x_{\mathrm{M}}(z)}
                                      \, \tfrac{1 + z'}{E(z')} \, \D z' \;,
\end{multline}
where we have used the dimensionless distance and Hubble functions.  The
integral in~\eqref{eq:kappa} presents two immediate problems for our
computations:  Firstly, we do not have access to the continuous matter
distribution~$\delta$, but only the discretised matter fields~$\delta_i$ in each
shell.  And secondly, the integral in~\eqref{eq:kappa} depends on all matter
below the source redshift~$z$, while we want to perform the computation
iteratively, keeping only a limited number of matter fields in memory.

To solve these problems, we impose three additional requirements for the matter
shells~$i = 0, 1, \ldots$ and their matter weight functions~$W_i$.  The first
requirement is that every shell~$i$ has an associated effective
redshift~$\bar{z}_i$ which is, in some sense, representative of the shell.  For
example, this could be the mean redshift of the matter weight function,
\begin{equation}
\label{eq:zeff}
    \bar{z}_i
    = \frac{\int z \, W_i(z) \, \D z}{\int W_i(z) \, \D z} \;,
\end{equation}
but other reasonable choices exist.  The second requirement is that the matter
weight functions of shells~$j < i$ vanish beyond the effective
redshift~$\bar{z}_i$,
\begin{equation}
\label{eq:w-req-2}
    W_j(z) = 0 \qquad \text{($j < i$ and $z \ge \bar{z}_i$)} \;.
\end{equation}
The third requirement is that the matter weight functions of shells~$j > i$
vanish below the effective redshift~$\bar{z}_i$,
\begin{equation}
\label{eq:w-req-3}
    W_j(z) = 0 \qquad \text{($j > i$ and $z \le \bar{z}_i$)} \;.
\end{equation}
In short, the requirements say that each matter shell has a representative
redshift which partitions the matter weight functions of all other shells.  This
is clearly the case for the effective redshifts~\eqref{eq:zeff} and the matter
weight function~\eqref{eq:w-redshift}.

To then approximate the continuous integral~\eqref{eq:kappa} by a discrete sum,
we first have to bring the integrand into a shape that matches the
definition~\eqref{eq:delta_i} of the integrated matter fields.  Using the
trivial partition of unity
\begin{equation}
    \sum_{j} \frac{W_j(z)}{\sum_{k} W_k(z)} = 1 \;,
\end{equation}
where the sums extend over all shells, we can introduce the matter weight
function~$W_i$ into the convergence~\eqref{eq:kappa},
\begin{multline}
\label{eq:kappa:1}
    \kappa(\n; z) \\
    = \tfrac{3\Omega_m}{2} \sum_{j}
        \int_{0}^{z} \! \delta\bigl(d_{\mathrm{c}}(z') \, \n\bigr) \, W_j(z') \,
                        q(z'; z) \, \D z' \;,
\end{multline}
with the function~$q$ being short for the geometric and weight factors,
\begin{equation}
\label{eq:qlens}
    q(z'; z)
    = \frac{1}{\sum_{k} W_k(z')} \,
        \frac{x_{\mathrm{M}}(z') \, x_{\mathrm{M}}(z', z)}{x_{\mathrm{M}}(z)} \,
        \frac{1 + z'}{E(z')} \;.
\end{equation}
To make our approximation, we now assume that the weight function~$W_j$ in the
integral~\eqref{eq:kappa:1} is so localised that the function~$q$ is constant
and equal to its value at the effective redshift~$\bar{z}_j$ for shell~$j$,
\begin{multline}
\label{eq:kappa:2}
    \kappa(\n; z) \\
    \approx \tfrac{3\Omega_m}{2} \sum_{j} q(\bar{z}_j; z)
        \int_{0}^{z} \! \delta\bigl(d_{\mathrm{c}}(z') \, \n\bigr) \, W_j(z') \,
                        \D z' \;.
\end{multline}
If the support of~$W_j$ corresponds to a thin shell, this holds for~$z >
\bar{z}_j$ as long the sum of weights in~\eqref{eq:qlens} changes as slowly as
the cosmological quantities.  We can then evaluate the
convergence~\eqref{eq:kappa:2} in the effective redshift~$\bar{z}_i$ for a given
shell~$i$:  By requirement~\eqref{eq:w-req-2}, we can truncate the sum before
shell~$i$, since~$q(\bar{z}_i; \bar{z}_i) = 0$ by definition,
\begin{multline}
\label{eq:kappa:3}
    \kappa_i(\n) = \kappa(\n; \bar{z}_i) \\
    = \tfrac{3\Omega_m}{2}
        \sum_{j=0}^{i-1} q(\bar{z}_j; \bar{z}_i)
        \int_{0}^{\bar{z}_i} \! \delta\bigl(d_{\mathrm{c}}(z') \, \n\bigr) \,
                                W_j(z') \, \D z' \;,
\end{multline}
and by requirement~\eqref{eq:w-req-3}, we can extend the remaining integrals
over all redshifts.  If we compare the resulting expression and the integrated
matter fields~\eqref{eq:delta_i}, we find that we can indeed write a discrete
approximation of the convergence,
\begin{equation}
\label{eq:kappa-approx}
    \kappa_i(\n)
    = \tfrac{3\Omega_m}{2} \sum_{j=0}^{i-1}
        \tfrac{x_{\mathrm{M}}(\bar{z}_j) \, x_{\mathrm{M}}(\bar{z}_j, \bar{z}_i)}
              {x_{\mathrm{M}}(\bar{z}_i)} \,
        \tfrac{1 + \bar{z}_j}{E(\bar{z}_j)} \, w_j \, \delta_j(\n) \;,
\end{equation}
where we have defined the lensing weights~$w_j$ to contain the dependency on the
matter weight functions,\footnote{%
The sum over weights in~\eqref{eq:qlens} reduces to a single term because of the
requirements~\eqref{eq:w-req-2} and~\eqref{eq:w-req-3} on the matter weight
functions in the effective redshift~$\bar{z}_j$.
}
\begin{equation}
\label{eq:w-lens}
    w_j
    = \frac{1}{W_j(\bar{z}_j)} \int W_j(z) \, \D z \;.
\end{equation}
The approximation~\eqref{eq:kappa-approx} as such is well known: Lensing can be
approximated by collapsing a continuous matter distribution onto a set of
discrete lensing planes.   Our main insight here is the exact form of the
lensing weights~\eqref{eq:w-lens} for the given matter weight functions, as well
as the requirements~\eqref{eq:w-req-2} and~\eqref{eq:w-req-3} on them.

Although the convergence~\eqref{eq:kappa-approx} is now discretised, it still
cannot be computed iteratively, since the geometric factor in each term depends
explicitly on the shells~$i$ and~$j$.  Here, the distance ratio relation of
\citet{2016A&A...592L...6S} is a powerful tool:  For~$i \ge 2$, define the ratio
of distance ratios
\begin{equation}
    t_i
    = \frac{x_{\mathrm{M}}(\bar{z}_{i-2}, \bar{z}_i)}
           {x_{\mathrm{M}}(\bar{z}_i)}
    \bigg/
    \frac{x_{\mathrm{M}}(\bar{z}_{i-2}, \bar{z}_{i-1})}
         {x_{\mathrm{M}}(\bar{z}_{i-1})} \;.
\end{equation}
The distance ratios for any other redshift~$\bar{z}_j$ then obey
\begin{equation}
\label{eq:drr}
    \frac{x_{\mathrm{M}}(\bar{z}_j, \bar{z}_i)}{x_{\mathrm{M}}(\bar{z}_i)}
    = t_i \, \tfrac{x_{\mathrm{M}}(\bar{z}_j, \bar{z}_{i-1})}
                   {x_{\mathrm{M}}(\bar{z}_{i-1})}
    + (1 - t_i) \, \tfrac{x_{\mathrm{M}}(\bar{z}_j, \bar{z}_{i-2})}
                         {x_{\mathrm{M}}(\bar{z}_{i-2})} \;.
\end{equation}
As shown by \citet{2016A&A...592L...6S}, this relation is exact and a
consequence of the mathematical form of the transverse comoving distance in
generic Robertson-Walker space-times.  Inserting~\eqref{eq:drr} into the
discrete approximation~\eqref{eq:kappa-approx}, we immediately obtain a
recurrence relation for the convergence,
\begin{multline}
\label{eq:kappa-rec}
    \kappa_i(\n)
    = t_i \, \kappa_{i-1}(\n) + (1 - t_i) \, \kappa_{i-2}(\n) \\
    + \tfrac{3\Omega_m}{2} \,
        \tfrac{x_{\mathrm{M}}(\bar{z}_{i-1}) \,
                x_{\mathrm{M}}(\bar{z}_{i-1}, \bar{z}_i)}
              {x_{\mathrm{M}}(\bar{z}_i)} \,
        \tfrac{1 + \bar{z}_{i-1}}{E(\bar{z}_{i-1})} \,
        w_{i-1} \, \delta_{i-1}(\n) \;.
\end{multline}
This is equivalent to the multi-plane formalism for the deflection in strong
gravitational lensing \citep{2014MNRAS.445.1954P, 2019A&A...624A..54S}.

\begin{figure}%
\centering%
\includegraphics[scale=0.85]{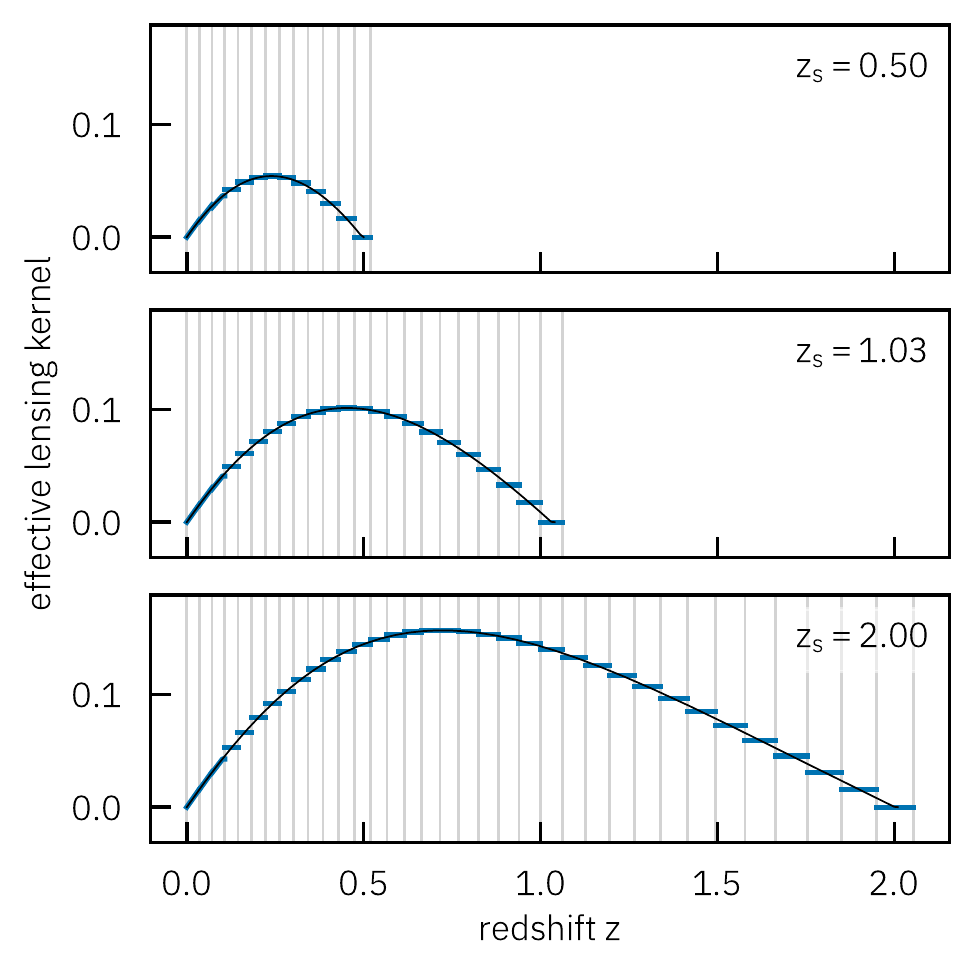}%
\caption{%
Effective lensing kernel of the lensing recurrence for source
redshifts~$z_{\mathrm{s}} = 0.50$ (\emph{top}), $z_{\mathrm{s}} = 1.03$
(\emph{middle}), and~$z_{\mathrm{s}} = 2.00$ (\emph{bottom}).  Vertical lines
indicate the boundaries of matter shells with constant thickness in comoving
distance $\Delta d_{\mathrm{c}} = 150$~Mpc.  Also shown is the true lensing
kernel (\emph{black}).
}%
\label{fig:lensing_kernel}%
\end{figure}

\begin{figure}%
\centering%
\includegraphics[scale=0.85]{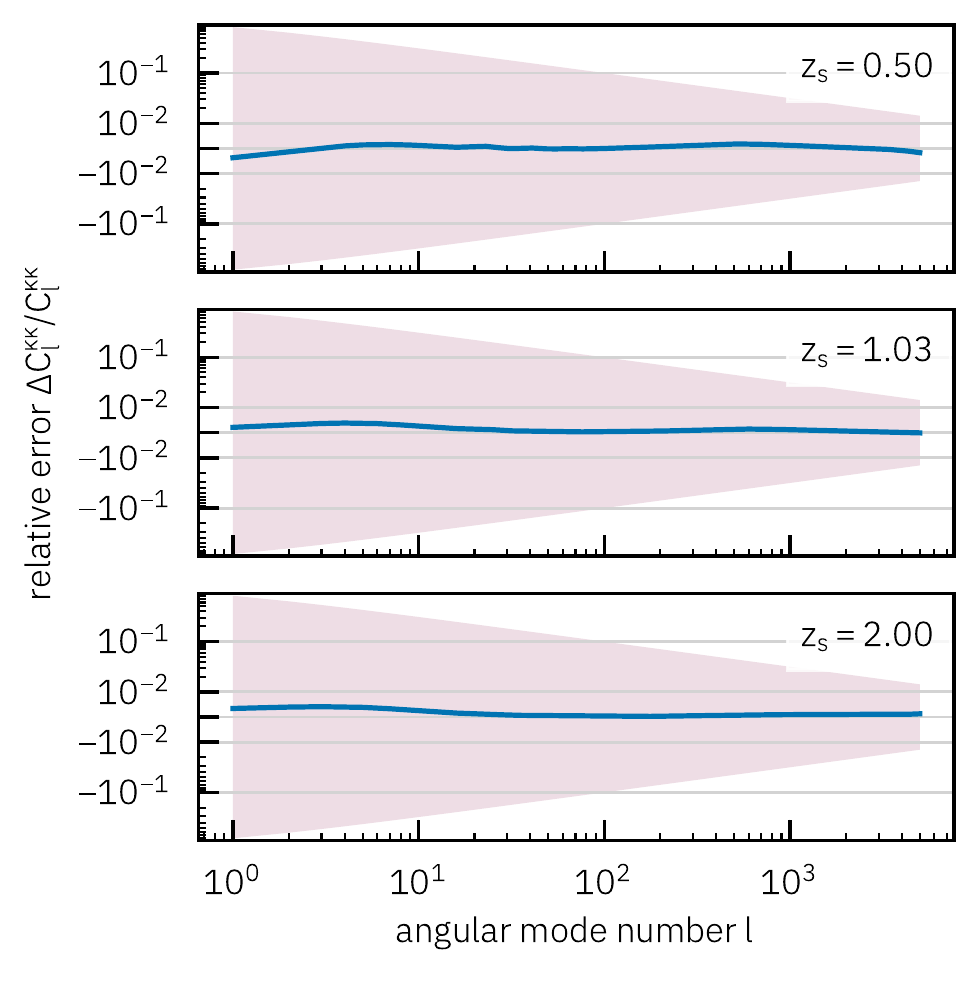}%
\caption{%
The relative error in the angular power spectra for the effective lensing
kernels of the lensing recurrence, as shown in Figure~\ref{fig:lensing_kernel}.
The shaded area shows the standard deviation of a Gaussian field for comparison.
Here and below, the logarithmic $y$-axis changes to linear when passing through
the origin.
}%
\label{fig:lensing_approx}%
\end{figure}

Overall, we have obtained the lensing recurrence~\eqref{eq:kappa-rec} by making
specific choices for our matter weight functions, and one single approximation
in~\eqref{eq:kappa:2}.  To test this approximation, we can compare the effective
lensing kernel of the recurrence, i.e.\ the resulting factor in~\eqref{eq:kappa}
multiplying~$\delta$, to the true lensing kernel.  This is done in
Figure~\ref{fig:lensing_kernel} for source redshifts~$z_{\mathrm{s}} = 0.50$,
$z_{\mathrm{s}} = 1.03$, and~$z_{\mathrm{s}} = 2.00$.  For the matter shells, we
use a constant size of~$\Delta d_{\mathrm{c}} = 150$~Mpc in comoving distance,
which is a reasonable choice, as we show in Section~\ref{sec:simulation}.  The
effective lensing kernel of our approximation is essentially the matter weight
function in each shell, scaled by the lensing recurrence, so that the flat
matter weight function~\eqref{eq:w-redshift} is a good global approximation to
the true kernel.  As one would expect, thinner shells result in a better
approximation, since we are essentially computing the convergence
integral~\eqref{eq:kappa} as a Riemann sum.  For the same reason, the
approximation improves naturally with higher source redshifts, which cover a
larger number of shells.

For a more quantitative check, we can compute the angular power spectra of the
effective lensing kernels, and compare the results to the true angular
convergence power spectra for each source redshift.  We compute the true spectra
with \emph{CAMB}, for angular modes up to number~$l = 5\,000$, without Limber's
approximation.  Figure~\ref{fig:lensing_approx} shows the resulting relative
errors.  For shells with~$\Delta d_{\mathrm{c}} = 150$~Mpc, the error is well
below the per cent level, and much smaller than the expected uncertainty due to
cosmic variance, which we approximate here by the Gaussian one for the sake of
simplicity.

\subsection{Shear}

Having found the convergence~\eqref{eq:kappa} for weak lensing by our simulated
matter distribution, we can obtain other weak lensing fields by applying the
spin-raising and spin-lowering operators~$\eth$ and~$\bar{\eth}$ \citep[see
e.g.][]{2016JMP....57i2504B}.  Their effect on the spin-weighted spherical
harmonics~${}_sY_{lm}$ is
\begin{align}
\label{eq:spup}
    \eth \, {}_sY_{lm}
    &= +\sqrt{(l-s)(l+s+1)} \, {}_{s+1}Y_{lm} \;,
\\
\label{eq:spdn}
    \bar{\eth} \, {}_sY_{lm}
    &= -\sqrt{(l+s)(l-s+1)} \, {}_{s-1}Y_{lm} \;,
\end{align}
where the spin-$0$ spherical harmonic~${}_0Y_{lm}$ is the scalar spherical
harmonic~$Y_{lm}$.

On the sphere, the Poisson equation for weak lensing reads
\begin{equation}
\label{eq:wl-poiseq}
    2\kappa = \eth \bar{\eth} \psi \;,
\end{equation}
and relates the convergence~$\kappa$ to the lensing (or deflection)
potential~$\psi$.  Let~$\kappa_{lm}$ be the modes of the spherical harmonic
expansion of the convergence field,
\begin{equation}
\label{eq:kappa_lm}
    \kappa(\n)
    = \sum_{lm} \kappa_{lm} \, Y_{lm}(\n) \;,
\end{equation}
and similarly~$\psi_{lm}$ for the lensing potential~$\psi$.  Inserting the
expansions into~\eqref{eq:wl-poiseq} and applying the operators~\eqref{eq:spup}
and~\eqref{eq:spdn}, the Poisson equation in harmonic space reduces to a simple
algebraic relation between the modes~$\kappa_{lm}$ and~$\psi_{lm}$,
\begin{equation}
\label{eq:kappa-psi}
    2 \kappa_{lm}
    = -l \, (l+1) \, \psi_{lm} \;.
\end{equation}
We can readily solve for~$\psi_{lm}$, except when~$l = m = 0$.  The
mode~$\psi_{00}$, however, describes a constant offset of the potential without
physical meaning, and can be given an arbitrary value.  We can thus completely
determine the lensing potential from the convergence via the spherical harmonic
expansion.

The principal observational effect of weak gravitational lensing, discussed
below in Section~\ref{sec:galaxies}, is caused by the shear field, commonly
denoted~$\gamma$.  Shear is the spin-$2$ field obtained by applying~$\eth$ twice
to the lensing potential,
\begin{equation}
    2 \gamma = \eth \eth \psi \;.
\end{equation}
As before, we can obtain an algebraic relation between the modes~$\gamma_{lm}$
of the shear field and~$\psi_{lm}$,
\begin{equation}
\label{eq:gamma-psi}
    2 \gamma_{lm}
    = \sqrt{(l+2) \, (l+1) \, l \, (l-1)} \, \psi_{lm} \;.
\end{equation}
An alternative definition is sometimes used where the shear is a spin-$(-2)$
field~$\gamma = \bar{\eth} \bar{\eth} \psi$.  However, this yields exactly the
same modes~\eqref{eq:gamma-psi}.  The difference between the definitions is
whether the coordinate system is left- or right-handed, and the shear in one
definition is the complex conjugate of the shear in the other.

From~\eqref{eq:gamma-psi}, it follows that the shear modes with~$l < 2$ vanish
identically, as expected for a spin-$2$ field.  We can hence treat the
case~$\gamma_{00} = 0$ separately, and compute the remaining shear modes with~$l
> 0$ directly from the convergence modes by combining~\eqref{eq:gamma-psi}
and~\eqref{eq:kappa-psi},
\begin{equation}
\label{eq:gamma-kappa}
    \gamma_{lm}
    = -\sqrt{\frac{(l+2) \, (l-1)}{l \, (l+1)}} \, \kappa_{lm} \;.
\end{equation}
While this implies that the difference between the modes of convergence and
shear vanishes for large~$l$, it is as much as~18\% at $l = 2$, so that the
conversion factor in~\eqref{eq:gamma-kappa} should always be applied.

In practice, we can hence construct a map of the shear field~$\gamma$ as
follows: Compute the discrete spherical harmonic transform~\eqref{eq:kappa_lm}
from a map of the convergence field, convert from convergence to shear
using~\eqref{eq:gamma-kappa}, and compute the inverse discrete spherical
harmonic transform.  This can once again be efficiently done using
\emph{HEALPix}.  We thus obtain maps of the shear field at the discrete source
redshifts of the convergence maps.

\section{Galaxies}
\label{sec:galaxies}

So far, we have developed robust methods to simulate the matter and weak lensing
fields, but neither of these are directly accessible to observations.  For that,
we need galaxies, which are tracers of both the matter field (through the
clustering of their positions), and of the weak lensing field (through the
distortion of their observed shapes).

Positions and shapes of galaxies are thus the fundamental observables for
cosmological galaxy surveys, and we must simulate them.  We have seen that the
weak lensing fields depend on the redshift of a given source, and we must hence
also assign redshifts to our simulated galaxies.  We may also wish to emulate
the tomographic binning of galaxies along the line of sight, which is typical of
modern galaxy surveys for weak lensing.  In wide-field surveys, this is usually
not done using the true, or at least spectroscopically-measured, redshift, but a
photometric redshift estimate, and this additional source of uncertainty should
be taken into account as well.  Besides, there are not only observational, but
also astrophysical effects which subtly change the expected clustering or weak
lensing signal of galaxies, such as their intrinsic alignment due to the
influence of a common tidal field from the large-scale structure of the
universe.

While all of these are complex phenomena in their own right, the fact that we
are merely using galaxies as tracers of other, hidden observables works greatly
in our favour.  After all, if we are not interested in e.g.\ the shapes of
galaxies as such, but only in what they can tell us about the two-point
statistics of the weak lensing fields, then it suffices to pick a simple model
of the former, as long as it accurately reproduces the latter.

In this section, we will therefore not spend too much time on specific models of
galaxy properties, but describe in rather general terms how individual models
can be combined into a whole simulation.

\subsection{Galaxy positions}

To sample galaxy positions in a given shell~$i$, we start by constructing the
\emph{HEALPix} map of galaxy number counts~$N^{\mathrm{g}}_{i,k}$.  We
parametrise~$N^{\mathrm{g}}_{i,k}$ in a manner that is similar to the matter
density,
\begin{equation}
\label{eq:gal-dens}
    N^{\mathrm{g}}_{i,k}
    = \bar{N}^{\mathrm{g}}_{i,k} \, [1 + \delta^{\mathrm{g}}_{i,k}] \;,
\end{equation}
where~$\bar{N}^{\mathrm{g}}_i$ the mean galaxy number in each \emph{HEALPix}
pixel, and~$\delta^{\mathrm{g}}_{i,k}$ is a \emph{HEALPix} map of the
discretised galaxy density contrast.  While~$\bar{N}^{\mathrm{g}}_i$ is a free
parameter of the simulated survey, the galaxy density
contrast~$\delta^{\mathrm{g}}_i$ must trace the realised large-scale structure
of the simulation.  We therefore express~$\delta^{\mathrm{g}}_i$ as a function
of the projected matter density contrast~$\delta_i$ of the shell using a generic
galaxy bias model~$B_{\mathrm{g}}$,
\begin{equation}
\label{eq:delta_g}
    \delta^{\mathrm{g}}_{i,k}
    = B_{\mathrm{g}}\bigl(\delta_{i,k}\bigr) \;.
\end{equation}
The bias function~$B_{\mathrm{g}}$ can in principle be arbitrarily complicated,
and depend not only on~$\delta_{i,k}$ but also explicitly on e.g.\ position,
redshift, or tidal field \citep[see e.g.][]{2018PhR...733....1D}.\footnote{%
Since~$\delta_i$ is the discretised field, any non-linear bias model will
also implicitly depend somewhat on the chosen shell boundaries, matter weight
functions, and resolution of the maps.
}

The most common choice of bias model is a linear bias~$\delta_{\mathrm{g}} =
b(z) \, \delta$, where~$b(z)$ is a redshift-dependent bias parameter.  On linear
scales, such a model is accurate and well-motivated;  besides, it makes
theoretical computation of the angular galaxy power spectra relatively
straightforward.  Because we apply the bias model~\eqref{eq:delta_g} to the
integrated matter fields~\eqref{eq:delta_i} in shells, we must translate a
continuous redshift-dependent bias parameter~$b(z)$ into an effective bias
parameter~$b_i$ for shell~$i$.  For that, we use a weighted mean,
\begin{equation}
\label{eq:b_i}
    b_i
    = \frac{\int b(z) \, W_i(z) \, \D z}{\int W_i(z) \, \D z} \;,
\end{equation}
where~$W_i$ is the matter weight function.  The typical shell size in redshift
of our simulations is~$\Delta z \lesssim 0.1$, so that the effective
bias~\eqref{eq:b_i} is usually a good approximation.

Having obtained the galaxy number counts~\eqref{eq:gal-dens} from the matter
field and a bias model, we can further adjust the resulting full-sky
map~$N^{\mathrm{g}}_{i,k}$ to account for observational details such as e.g.\
the survey footprint or varying survey depth.  We describe these effects using
an optional visibility map:  Each number~$N^{\mathrm{g}}_{i,k}$ is multiplied by
a visibility value~$V_{i,k}$ between~$0$ and~$1$ that is the probability of
observing a galaxy in \emph{HEALPix} pixel~$k$ for shell~$i$.

With the final map of expected galaxy numbers~$N^{\mathrm{g}}_{i,k}$
constructed, we sample the realised number of galaxies in each \emph{HEALPix}
pixel from some given distribution.  The Poisson distribution is commonly
assumed, but any other choice is possible.  Finally, we pick for each galaxy a
uniformly random position inside its \emph{HEALPix} pixel.  Overall, we thus
obtain an observed galaxy distribution that traces the large-scale structure of
our simulation.

\begin{figure}%
\centering%
\includegraphics[scale=0.85]{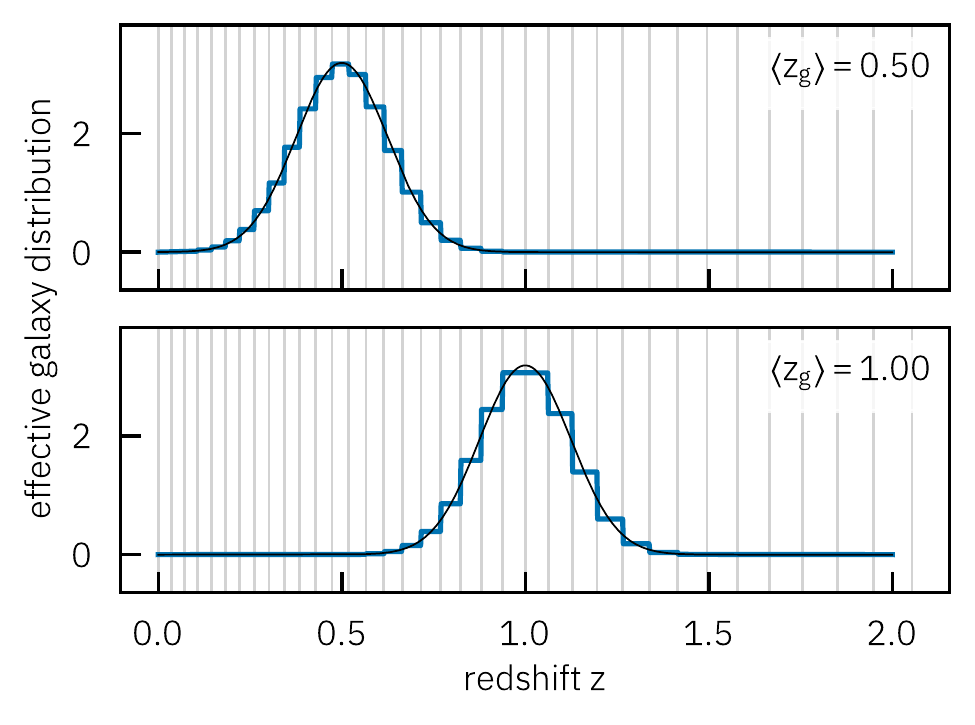}%
\caption{%
Effective redshift distribution (\emph{blue}) of the discretised galaxy density
contrast~$\delta^{\mathrm{g}}_i$ for two representative populations
(\emph{black}) with mean redshifts~$\ev{z_{\mathrm{g}}} = 0.5$ (\emph{top})
and~$\ev{z_{\mathrm{g}}} = 1.0$ (\emph{bottom}).  Vertical lines indicate the
boundaries of matter shells with constant thickness in comoving distance $\Delta
d_{\mathrm{c}} = 150$~Mpc.
}%
\label{fig:galaxies_kernel}%
\end{figure}

\begin{figure}%
\centering%
\includegraphics[scale=0.85]{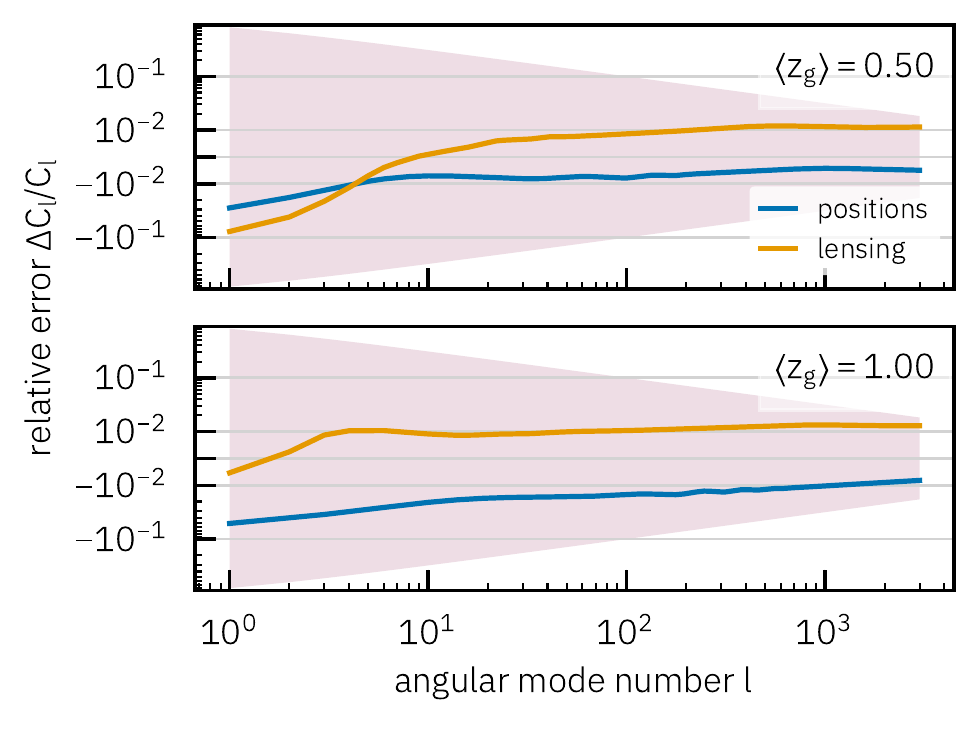}%
\caption{%
The relative error in the angular power spectra for the effective galaxy
distribution of the discretised galaxy field, as shown in
Figure~\ref{fig:galaxies_kernel}, for positions (\emph{blue}) and lensing
(\emph{orange}).  The shaded area shows the standard deviation of a Gaussian
field for comparison.
}%
\label{fig:galaxies_approx}%
\end{figure}

Because we sample galaxy positions from the discretised galaxy density
contrast~$\delta^{\mathrm{g}}_i$, all galaxies in a given shell~$i$ follow the
same matter density field~$\delta_i$, given by the
projection~\eqref{eq:delta_i}.  As far as the two-point statistics are
concerned, the effective redshift distribution of the galaxies in shell~$i$ is
therefore determined by the matter weight function~$W_i$.  This is shown in
Figure~\ref{fig:galaxies_kernel} for shells of size~$\Delta d_{\mathrm{c}} =
150$~Mpc in comoving distance, and two representative Gaussian redshift
distributions with respective means~$z = 0.5$ and~$z = 1.0$ and the same
standard deviation~$\sigma_z = 0.125$.  Although the situation is ostensibly
similar to the lensing kernels in Figure~\ref{fig:lensing_kernel}, the smaller
size of the distributions compared to the shells results in relative errors at
the per cent level in the angular power spectra for the galaxy positions and
lensing, shown in Figure~\ref{fig:galaxies_approx}.  However, this level of
uncertainty in the galaxy distribution is comparable to that achieved by
observations \citep{2018PASJ...70S...9T, 2018AJ....155....1G,
2020A&A...644A..31E, 2021A&A...647A.124H, 2022MNRAS.511.2170C}, so that there is
little real incentive to push the errors down by decreasing the shell size.  In
fact, the observational uncertainty means that we can simply assume the
discretised distribution in Figure~\ref{fig:galaxies_kernel} to be the true
redshift distribution of our simulated survey, without introducing a significant
disagreement between simulations and observations.  If we apply this strategy,
errors from the discretisation of the matter fields disappear entirely in the
galaxies sector, for both angular clustering and weak lensing.

\subsection{Galaxy redshifts}

For the radial distribution of galaxies, we sample the true redshift~$z$ of
galaxies from a given redshift distribution $dN/dz$, with~$N$ the number
density of galaxies as a function of redshift.  This is done separately within
each matter shell.  Although the resulting galaxy redshifts will follow the
given distribution, they will not display any radial correlations on scales
smaller than the matter shells.  The choice of redshift distribution is
arbitrary, and could be the actual distribution from a galaxy survey, or the
commonly used distribution of \cite{1994MNRAS.270..245S} for photometric
surveys,
\begin{equation}
    p(z)
    \propto z^{\alpha} \, \E^{-(z/z_0)^{\beta}} \;,
\end{equation}
where $z_0$ is related to the median redshift of the distribution, while the
exponents~$\alpha$ and~$\beta$ are typically set to~2 and~1.5, respectively
\citep{2007MNRAS.381.1018A}.  We allow for multiple such redshift
distributions to be given, which might represent different samples or tracers of
large-scale structure.

We can additionally generate photometric galaxy redshifts~$z_{\mathrm{ph}} $ by
sampling from a conditional redshift distribution $p(z_{\rm ph} | z)$.  For
example, a redshift-dependent Gaussian error with standard deviation $\sigma(z)
= \sigma_0 \, (1 + z)$, parametrised by the error~$\sigma_0$ at $z = 0$, has the
conditional distribution
\begin{equation}
    p(z_{\mathrm{ph}} | z)
    = \frac{1}{\sigma(z) \, \sqrt{2\pi}} \exp\Bigl\{
        -\frac{1}{2} \, \Bigl(\frac{z_{\mathrm{ph}}-z}{\sigma(z)}\Bigr)^2
    \Bigr\} \;,
\end{equation}
which is readily numerically sampled.  If a more realistic and tailored
simulation is desired, any other conditional distribution can be used in place
of this simple model, such as e.g.\ the empirical photometric redshift
distribution of a given survey.

Finally, we note that both the true and the photometric redshift distributions
do not have to coincide at all with the matter shells, and can have arbitrary
overlaps.

\subsection{Galaxy shears}

One of the main cosmological observables in galaxy surveys is the shape of
objects.  It is quantified by the ellipticity~$\epsilon$, which is
complex-valued with components~$\epsilon_1$ and~$\epsilon_2$,
\begin{equation}
    \epsilon = \epsilon_1 + \I \, \epsilon_2 \;.
\end{equation}
The simplest case is the ellipticity of an elliptical isophote with axis
ratio~$q$, rotated by an angle~$\phi$ against the local coordinate frame,
\begin{equation}
\label{eq:eps-isophot}
    \epsilon = \frac{1 - q}{1 + q} \, \E^{2\I \phi} \;.
\end{equation}
For extended surface brightness distributions, the ellipticity is defined in
terms of the second moments of the distribution \citep[see
e.g.][]{2006glsw.conf.....M}.  It is strictly true that $|\epsilon| \le 1$,
which follows immediately from~\eqref{eq:eps-isophot} for an elliptical
isophote, and from positive definiteness of the second moments in the general
case.

The importance of the ellipticity~$\epsilon$ for cosmology is owed to the fact
that it is a tracer of the so-called reduced shear~$g$, which is a
complex-valued field that combines the convergence~$\kappa$ and shear~$\gamma$
from weak gravitational lensing,
\begin{equation}
\label{eq:g}
    g
    = \frac{\gamma}{1 - \kappa} \;.
\end{equation}
Under the influence of a reduced shear~$g$, the ellipticity~$\epsilon$ of a
small source transforms as
\begin{equation}
\label{eq:eps-wl}
    \epsilon \mapsto \frac{\epsilon + g}{1 + g^*\epsilon} \;.
\end{equation}
It was shown by \citet{1997A&A...318..687S} that if the unlensed galaxy
ellipticity distribution is isotropic, i.e.\ with no preferred direction, then
the expectation of the ellipticity~$\epsilon$ equals the reduced shear~$g$,
\begin{equation}
    \ev{\epsilon} = g \;.
\end{equation}
Although this result is often stated as an approximation to first order in~$g$
(which it is not), it holds exactly for any isotropic distribution of galaxy
ellipticities.  If we only care for galaxy ellipticities as tracers of the weak
lensing field, we thus have the freedom to choose any such distribution for our
simulation.

A common choice is to sample the ellipticity components~$\epsilon_1$
and~$\epsilon_2$ as independent normal random variates with a given standard
deviation~$\sigma_\epsilon$ in each component.  We present this model, as well
as a related but improved distribution, in Appendix~\ref{sec:app-c}.  For a more
realistic ellipticity distribution, we can sample the galaxy shape e.g.\ as a
triaxial ellipsoid under a random viewing angle \citep{2004ApJ...601..214R}.  In
this way, it is also possible to include even more subtle effects such as e.g.\
dust extinction and reddening, which depend on the viewing angle of the galaxy
\citep{2008MNRAS.388.1321P}.

For any chosen distribution, we sample an ellipticity for each galaxy in a given
shell~$i$.  We then interpolate the convergence map~$\kappa_i$ and shear
map~$\gamma_i$ at the galaxy position.  From these values, we compute the
reduced shear~\eqref{eq:g} and use the transformation law~\eqref{eq:eps-wl} to
give each galaxy an observed ellipticity under the effect of weak lensing.  As
commonly done, we call the weakly-lensed ellipticities the ``galaxy shears''.

\subsection{Intrinsic alignments}

Galaxies systematically align with the overall large-scale structure of the
universe \citep[for reviews, see][]{2015SSRv..193....1J, 2015SSRv..193...67K,
2015SSRv..193..139K}.  This effect breaks the assumed isotropy of the
distribution of galaxy shapes, and translates into correlations in the
ellipticities between physically close galaxies.  On the level of two-point
statistics, the result is a contamination of the cosmic shear signal by
so-called intrinsic alignments \citep{2000MNRAS.319..649H, 2002A&A...396..411K,
2003MNRAS.339..711H, 2007NJPh....9..444B}.

However, the fact that the signals from weak lensing and intrinsic alignments
are very similar can be exploited for simulations \citep{2019PASJ...71...43H,
2020MNRAS.498.4060G, 2021A&A...645A.104A, 2021MNRAS.501..954J}.  If we adjust
the convergence~$\kappa$ from weak lensing to include an effective
contribution~$\kappa^{\mathrm{IA}}$ from intrinsic alignments,
\begin{equation}
\label{eq:kappa_ia}
    \kappa \mapsto \kappa + \kappa^{\mathrm{IA}} \;,
\end{equation}
this is subsequently transformed into an effective shear
via~\eqref{eq:gamma-kappa}, and the resulting reduced shear~\eqref{eq:g}
imprints the correlation due to intrinsic alignments onto the isotropic galaxy
ellipticities at the same time as the shear.\footnote{%
We note that the effective~$\kappa^{\mathrm{IA}}$ is constructed under the
assumption of a linear relation between convergence, shear, and galaxy
ellipticity, which only holds to linear order; see~\eqref{eq:g}
and~\eqref{eq:eps-wl}.
}
To simulate intrinsic alignments in this manner, we add~$\kappa^{\mathrm{IA}}$
to our~$\kappa$ map before the galaxy ellipticities are sampled (but after all
simulation steps that require the true convergence have passed).

\begin{figure*}%
\centering%
\includegraphics[scale=0.85]{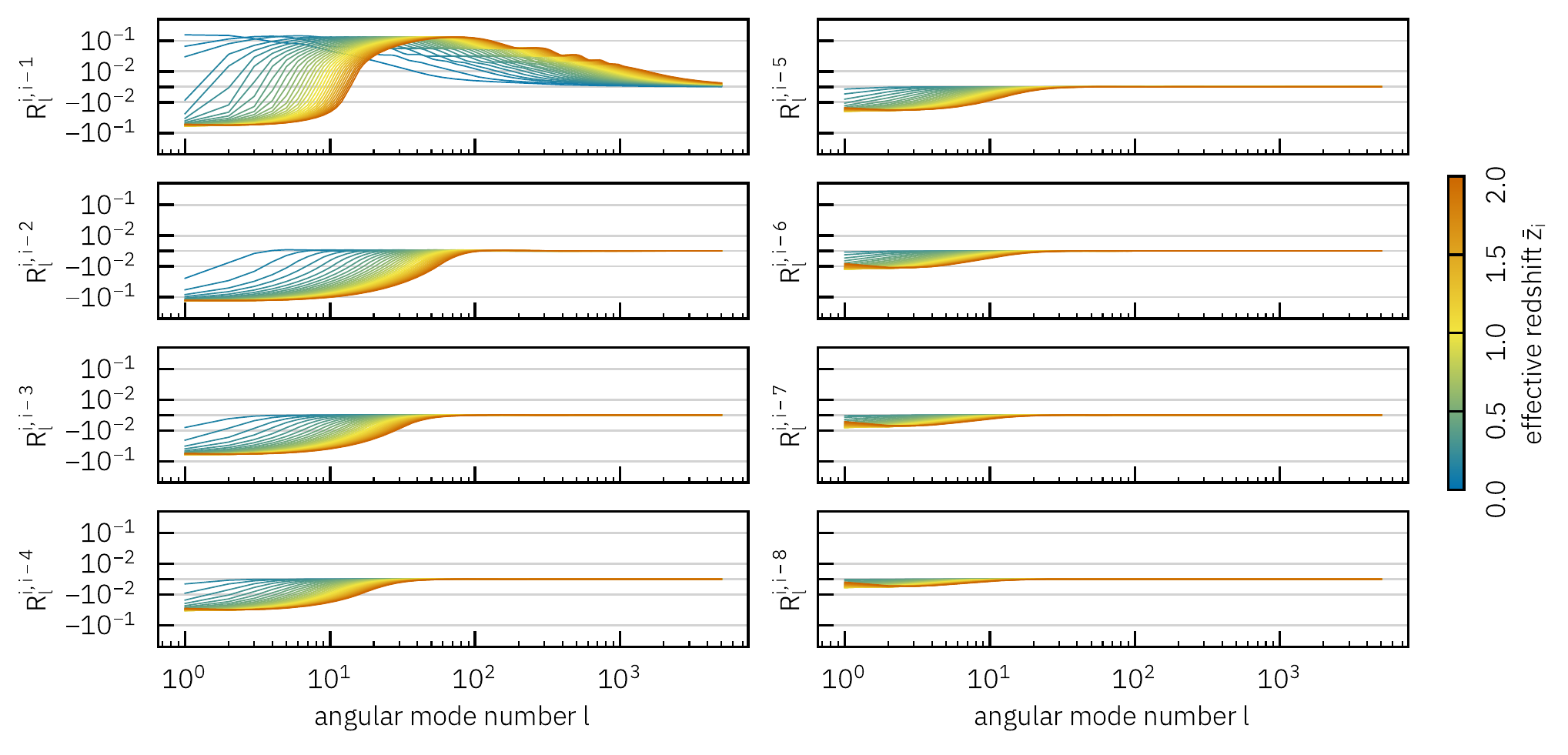}%
\caption{%
Correlation coefficient~$R_l^{ij}$ for the angular matter power spectrum of
shells with constant thickness $\Delta d_{\mathrm{c}} = 150$~Mpc in comoving
distance.  Shown are consecutive shells with $j - i = 1$ (\emph{top left}) to $j
- i = 8$ (\emph{bottom right}).  The colours indicate the effective redshift of
each shell from~$z=0$ to~$z=2$.
}%
\label{fig:correlations}%
\end{figure*}

To give a specific example, a widely used model to obtain the effective
convergence~\eqref{eq:kappa_ia} is the Non-Linear Alignment (NLA) model
\citep{2001MNRAS.320L...7C, 2004PhRvD..70f3526H, 2007NJPh....9..444B}.  It
proposes that the shear signal coming from intrinsic alignments is proportional
to the projected tidal field and hence ultimately to the matter density
contrast~$\delta$.  For a given shell~$i$, we compute the effective
contribution~$\kappa^{\mathrm{IA}}_i$ in~\eqref{eq:kappa_ia} from the projected
matter field~$\delta_i$,
\begin{multline}
    \kappa^{\mathrm{IA}}_i(\n) \\
    = -A_{\mathrm{IA}} \,
        \frac{C_1 \, \Omega_{\mathrm{m}} \,
            \overline{\rho}_{\mathrm{cr}}(\bar{z}_i)}
        {D(\bar{z}_i)} \,
        \bigg(\frac{1+\bar{z}_i}{1+z_{\mathrm{IA}}}\bigg)^{\eta(\bar{z}_i)} \,
                                                                \delta_i(\n) \;,
\end{multline}
where $A_{\mathrm{IA}}$ is the intrinsic alignment amplitude, $C_1$ is a
normalisation constant \citep{2004PhRvD..70f3526H},
$\overline{\rho}_{\mathrm{cr}}(\bar{z}_i)$ is the mean critical matter density
of the universe at a representative redshift $\bar{z}_i$ for shell~$i$,
$D(\bar{z}_i)$ is the growth factor normalised to unity today, and $\eta$ is the
index of a power law which describes the redshift dependence of the intrinsic
alignment strength relative to the tidal field with respect to the pivot
redshift~$z_{\mathrm{IA}}$.

\section{Simulating a weak lensing galaxy survey}
\label{sec:simulation}

We have implemented the simulation steps of the previous sections in a new,
publicly available computer code called \emph{GLASS}, the Generator for Large
Scale Structure.  In this section, we use \emph{GLASS} to demonstrate a
simulation that would be typical for a Stage~4 photometric weak lensing galaxy
survey such as \emph{Euclid}, \emph{LSST}, or \emph{Roman}.

Our initial Figure~\ref{fig:flowchart} provides a high-level flowchart for
how~\emph{GLASS} simulates individual shells.  In the matter sector, we specify
the shell boundaries and the matter weight functions, from which the angular
matter power spectra are computed.  For this example, we once again use
\emph{CAMB}, without Limber's approximation.  A lognormal matter field is
subsequently sampled from the angular matter power spectra, using a chosen
number of previous shells for correlations.

In the weak lensing sector, the matter weight functions are used to compute the
lensing weights~\eqref{eq:w-lens}.  The lensing weights and the matter field are
then used to iteratively compute the convergence field.  If intrinsic alignments
of galaxies are being simulated, their effect is added to the convergence field.
Finally, the shear field is computed from the convergence using a spherical
harmonic transform.

In the galaxies sector, the matter field is biased to sample the random galaxy
positions.  Galaxy redshifts are sampled directly from the provided source
distributions.  Galaxy ellipticities are sampled from a suitable distribution.
Positions and ellipticities then enter the computation of the galaxy shears: The
convergence and shear fields are interpolated using the galaxy redshifts and
evaluated at the galaxy positions to produce the reduced shears, which is
applied to the galaxy ellipticities to produce the final galaxy shears.

The outcome of these steps is a typical galaxy catalogue with positions,
redshifts, and shears, which can be used for what is often called ``3x2pt''
analysis.

We will now carry out a simulation to validate these results, which requires a
number of user choices.  The first is the distribution of the matter shell
boundaries, and hence the size of the shells.  Because the two-point statistics
of the matter field ultimately depend on physical distance, we generally choose
matter shells with a constant size in comoving distance.  As shown in
Sections~\ref{sec:lensing} and~\ref{sec:galaxies}, we obtain accurate results
from the respective approximations for lensing and galaxies when the matter
fields are discretised in shells of a constant size of~$\Delta d_{\mathrm{c}} =
150$~Mpc in comoving distance.  We therefore adopt this value here.

As explained in Section~\ref{sec:sampling}, we can then choose to only keep a
limited number of correlated matter shells in memory over the course of the
simulation, to reduce the computational burden imposed by such thin shells.  To
make an informed choice for said number, we quantify the correlation of the
matter fields between two shells~$i$ and~$j$ by introducing the correlation
coefficient~$R_l^{ij}$ of the angular matter power spectra,
\begin{equation}
    R_l^{ij}
    = \frac{C_l^{ij}}{\sqrt{C_l^{ii} \, C_l^{jj}}} \;.
\end{equation}
Angular power spectra are the (co)variances of the modes of the spherical
harmonic expansion, and~$R_l^{ij}$ is hence a proper correlation coefficient in
the usual sense:  It takes values between~$+1$ and~$-1$, with the former meaning
perfect correlation, and the latter meaning perfect anticorrelation.
Figure~\ref{fig:correlations} shows the correlation coefficient~$R_l^{ij}$ for
offsets~$j - i = 1, \ldots, 8$ in shells with~$\Delta d_{\mathrm{c}} = 150$~Mpc
at redshifts between~$0$ and~$2$.  We see how the correlation between shells
scans through the three-dimensional matter correlation function:  On scales
$\lesssim 150$~Mpc comoving, matter is largely positively correlated, which is
seen in adjacent shells.  This is compensated by negative correlation on larger
scales, which is seen in the non-neighbouring shells.

We now consult Figure~\ref{fig:correlations} to find the number of matter shells
to correlate.  If we wish to achieve per cent-level accuracy in the matter
sector at $l \approx 10$, say, we find that it suffices to keep five correlated
shells in memory over the entire redshift range, which is readily achievable on
standard computer hardware.  This level of accuracy is consistent with the
lensing sector, shown in Figure~\ref{fig:lensing_approx}, and the galaxies
sector, shown in Figure~\ref{fig:galaxies_approx}.  We can therefore make simple
and understandable choices about the simulation parameters, based on the desired
accuracy of the results.  Of course, the specific values we use depend entirely
on our adopted shell size of~$\Delta d_{\mathrm{c}} = 150$~Mpc.

\begin{figure}%
\centering%
\includegraphics[scale=0.85]{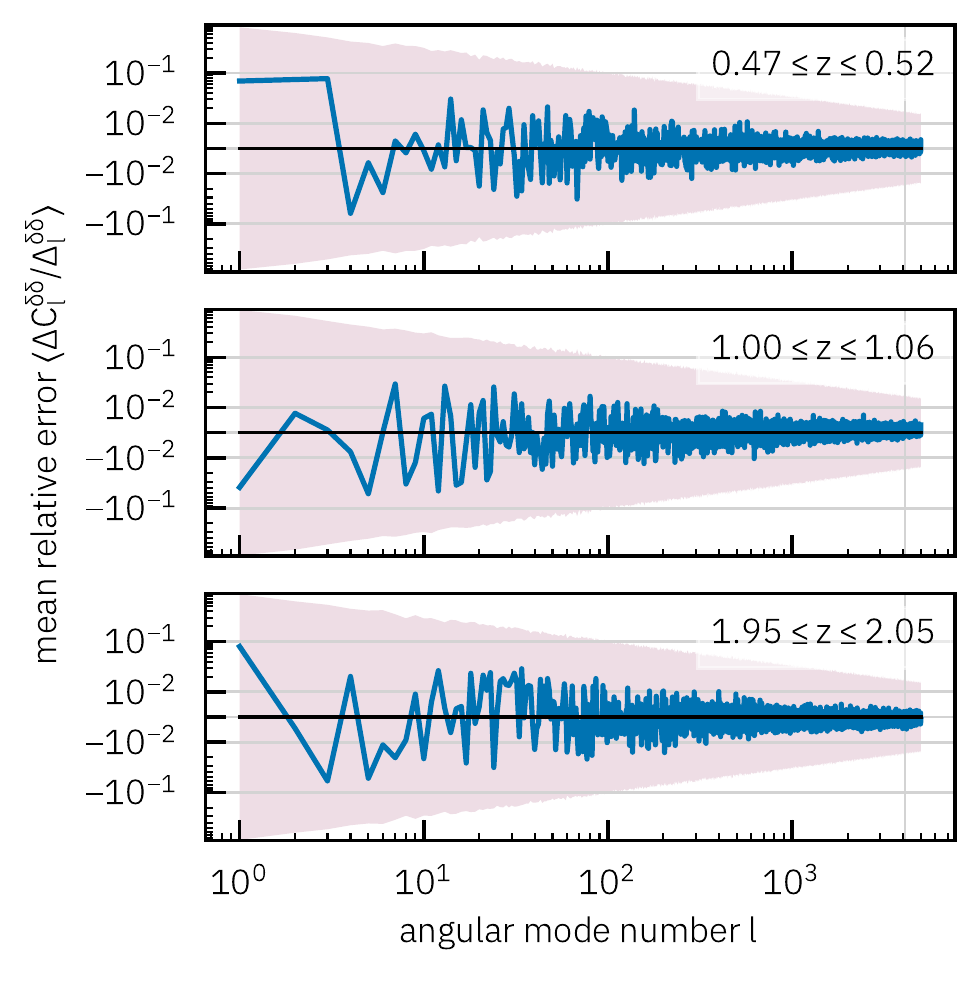}%
\caption{%
Mean relative error of the angular matter power spectra from~200 realisations of
a lognormal matter field.  Shown are three shells containing redshifts~$z$
with~$0.47 \le z \le 0.52$ (\emph{top}), $1.00 \le z \le 1.06$ (\emph{middle}),
and $1.95 \le z \le 2.05$ (\emph{bottom}).  The vertical line indicates
the~$N_{\mathrm{side}}$ parameter of the simulation.  The shaded area shows
cosmic variance of the realisations.
}%
\label{fig:delta_err}%
\end{figure}

To demonstrate that the realised matter field achieves our stated accuracy, we
create $200$ simulations of lognormal matter fields with angular modes up to~$l
= 5\,000$ from \emph{HEALPix} maps with~$N_{\mathrm{side}} = 4\,096$.
Figure~\ref{fig:delta_err} shows the mean relative error of the realised angular
matter power spectra for three representative shells with redshifts near~$z =
0.5$, $z = 1.0$, and $z = 2.0$.  The achieved error is well below the per cent
level, which in turn is well below the level of cosmic variance of the
realisations.  This level of accuracy in the recovered matter fields is not
currently attained by lognormal simulations \citep{2016MNRAS.459.3693X}, which
shows that our Gaussian angular power spectrum solver is working as intended.

\begin{figure}%
\centering%
\includegraphics[scale=0.85]{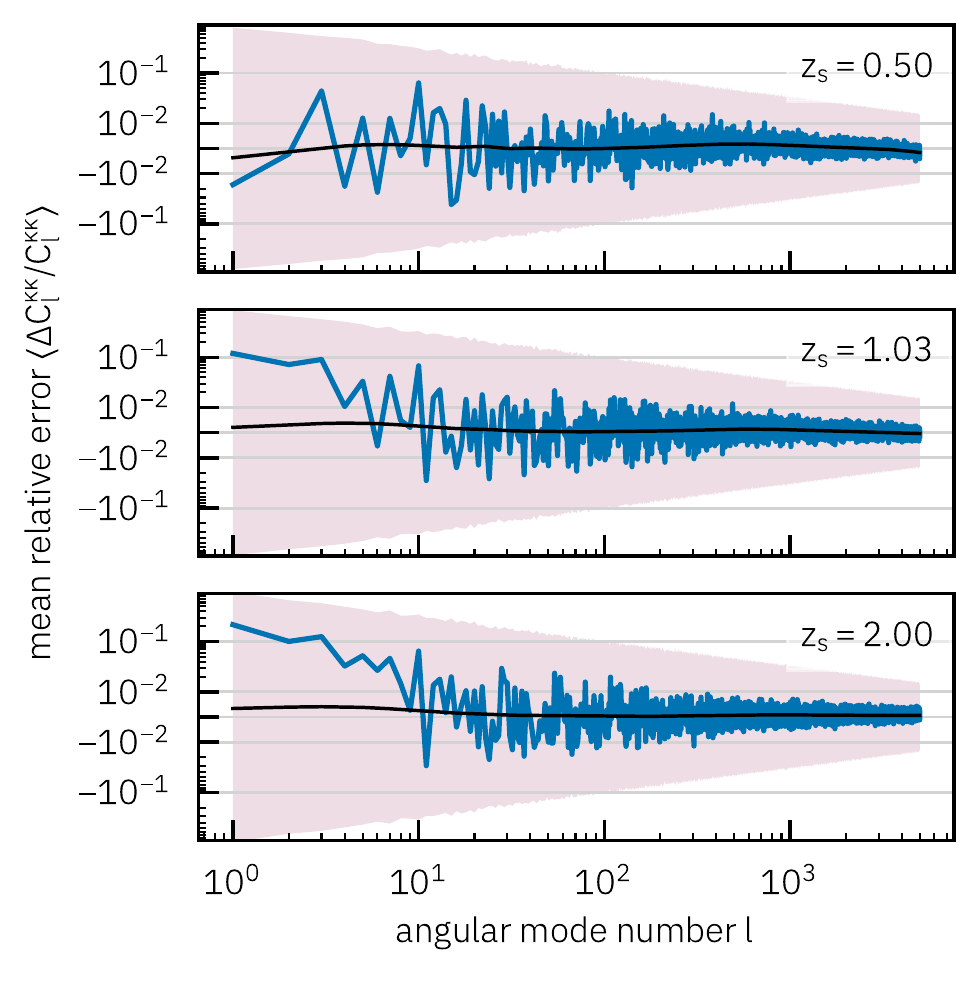}%
\caption{%
Mean relative error of the angular power spectra of the convergence from~200
realisations of a lognormal matter field with five correlated shells
(\emph{blue}).  Shown are source redshifts~$z_{\mathrm{s}} = 0.50$ (\emph{top}),
$z_{\mathrm{s}} = 1.03$ (\emph{middle}), and~$z_{\mathrm{s}} = 2.00$
(\emph{bottom}). Also shown is the expected curve from
Figure~\ref{fig:lensing_approx} for fully correlated shells (\emph{black}).  The
shaded area shows cosmic variance of the realisations.
}%
\label{fig:kappa_err}%
\end{figure}

Using the same $200$ realisations, we also demonstrate that the iterative
computation of the convergence field with the multi-plane
formalism~\eqref{eq:kappa-approx} achieves the desired accuracy.
Figure~\ref{fig:kappa_err} shows the mean relative error of the realised angular
power spectra for three source redshifts near~$z = 0.5$, $z = 1.0$, and $z =
2.0$.  The realisations agree with the theoretical predictions from
Figure~\ref{fig:lensing_approx} up to the point near $l \approx 10$ where
missing (anti-)correlations from the uncorrelated shells become significant,
according to Figure~\ref{fig:correlations}.  This missing negative correlation
explains why, for values of~$l \lesssim 10$, the simulated convergences have
angular power spectra which lie above the expectations.  Overall, the results of
Figure~\ref{fig:kappa_err} hence show not only that the multi-plane
approximation for weak lensing holds, but also that cross-correlations are
correctly imprinted on the matter fields.

As a final test, we simulate a catalogue of galaxies that is typical for
``3x2pt'' analysis with tomographic redshift bins.   Since we are only
interested in validation here, we use two redshift bins with small but not
insignificant overlap, which is the case where cross-correlations are most
difficult to get right.  In particular, we adopt the discretised distribution of
Figure~\ref{fig:galaxies_kernel} as the true galaxy distribution, so that we can
expect there to be no effect due to discretisation on our results.  We generate
$1\,000$ simulations of galaxy positions, redshifts, and shears, using a mean
number density of~1~galaxy per square arcminute in each tomographic bin.  To be
able to compute accurate theoretical predictions for the results, we use a
linear galaxy bias with constant bias parameter~$b = 0.8$.  This unrealistically
low value~$b < 1$ is necessary for accuracy of the theory, not our simulations:
If~$b > 1$, the galaxy density contrast~$\delta_{\mathrm{g}} = b \, \delta$ can
become less than~$-1$ in very underdense regions.  We would have to clip such
unphysical values to~$-1$ in our simulation, which effectively renders the model
non-linear, and deviates from the assumed theory.

\begin{figure*}%
\centering%
\includegraphics[scale=0.85]{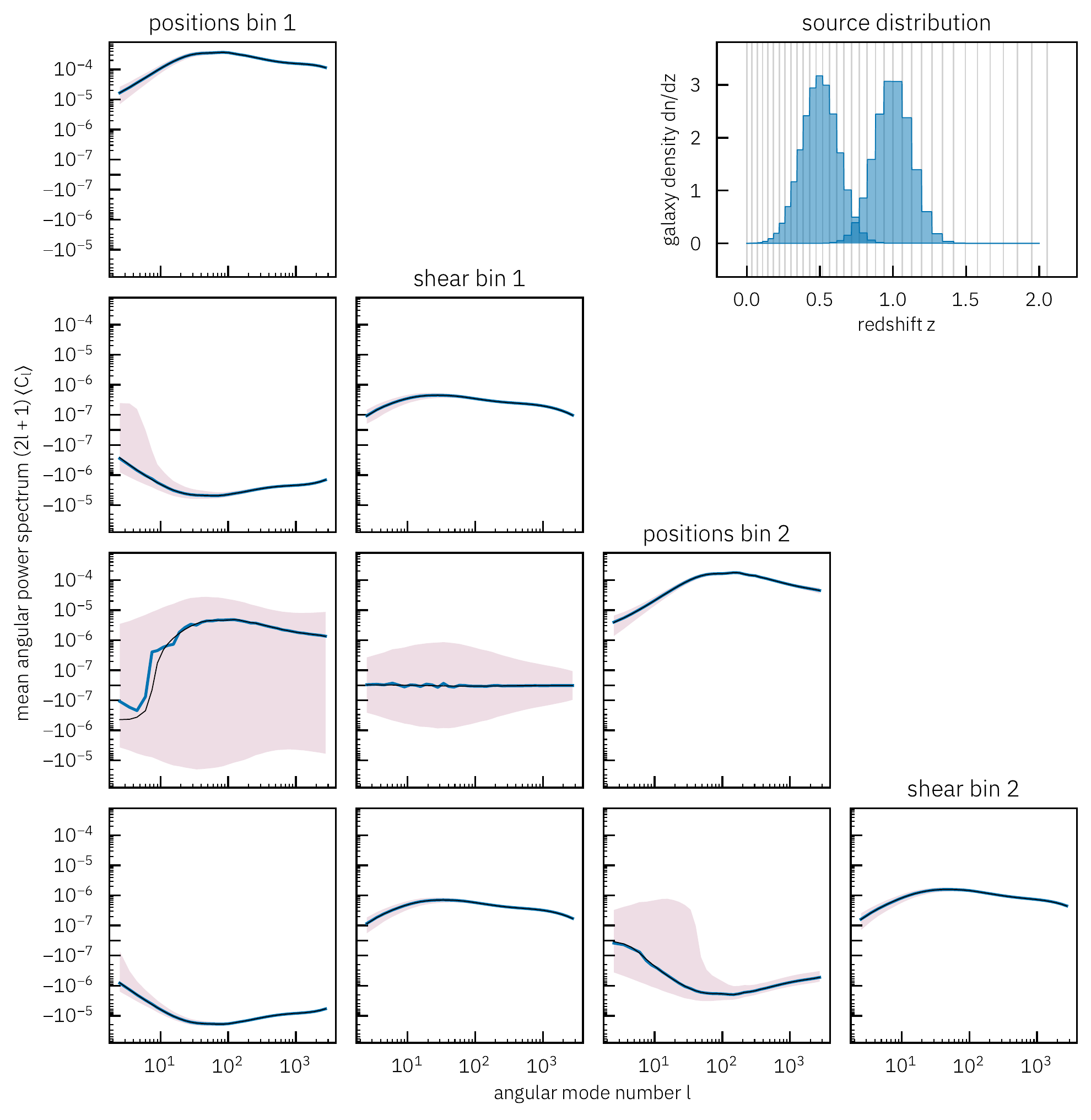}%
\caption{%
Mean angular power spectra (\emph{blue}) from $1\,000$ realisations of galaxy
positions and shears in a simulated full-sky survey.  Shown are the auto- and
cross-correlations for two tomographic redshift bins 1 and 2 with respective
mean redshifts of $\ev{z} = 0.5$ and $\ev{z} = 1.0$ (\emph{inset}).   Also shown
are the theoretical spectra computed by \emph{CAMB} (\emph{black}). To reduce
visible noise, the angular power spectra are averaged over 40 logarithmic bins
in angular mode number~$l$.  The shaded area shows cosmic variance of the
realisations.
}%
\label{fig:galaxies}%
\end{figure*}

\begin{figure*}%
\centering%
\includegraphics[scale=0.85]{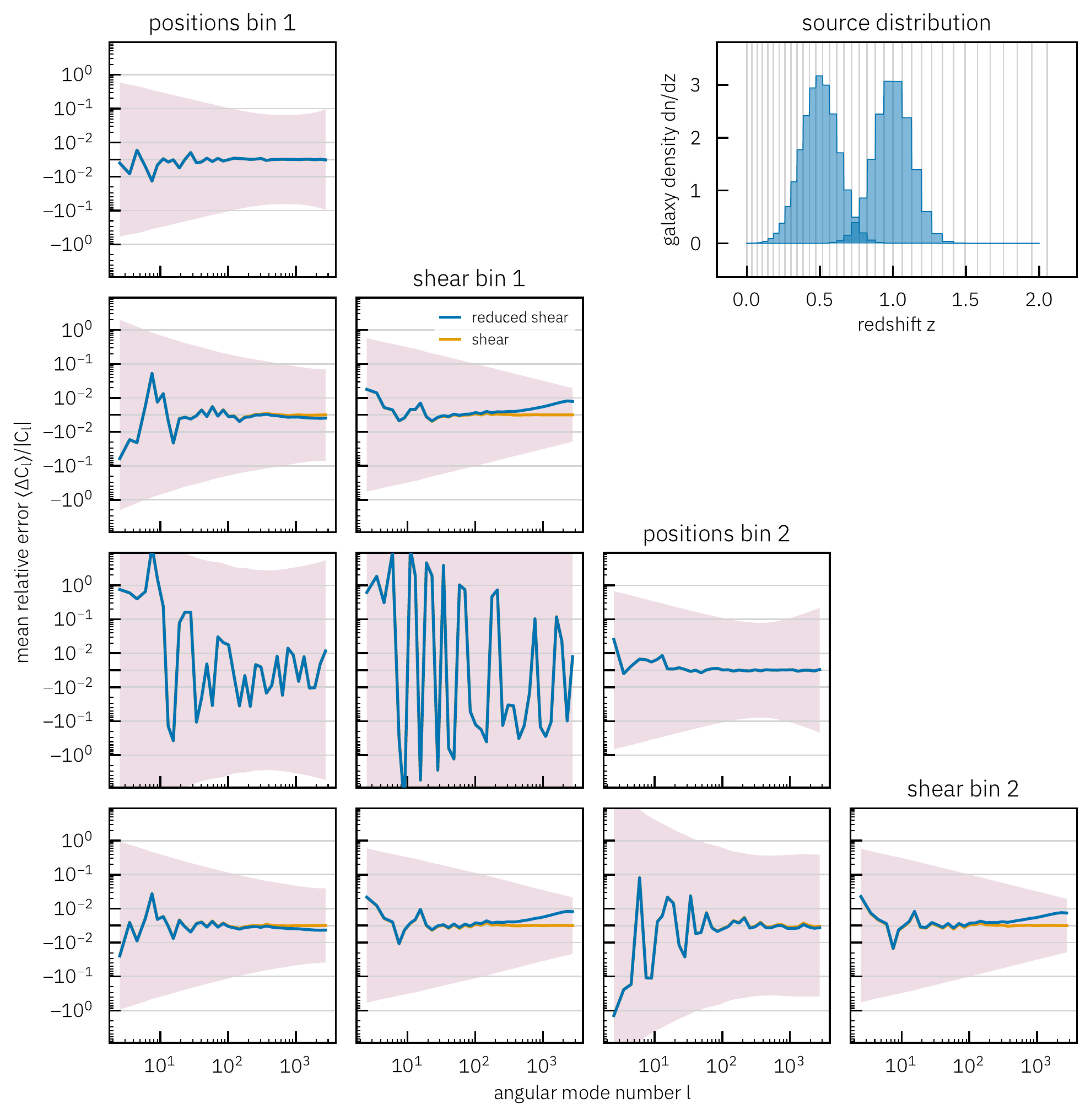}%
\caption{%
Mean relative error between the measured and theoretical angular power spectra
of Figure~\ref{fig:galaxies}.  Shown are results for both the reduced shear~$g$
(\emph{blue}), of which galaxies are a tracer, and the true shear~$\gamma$
(\emph{orange}), which is the fundamental gravitational lensing field.  The
theoretical spectra from \emph{CAMB} are computed for the latter instead of the
former.  The shaded area shows cosmic variance of the realisations.  The central
panel is essentially pure noise because there is only a vanishingly small
correlation between shear in the foreground and positions in the background.
}%
\label{fig:galaxies_err}%

\end{figure*}

For every combination of galaxy positions and shears across the two bins, we
then compare the realised angular power spectra to theory.  The results are
shown in Figures~\ref{fig:galaxies} and~\ref{fig:galaxies_err}.\footnote{%
The position--shear signal is sometimes shown with a positive sign when defined
as ``galaxy--galaxy lensing'' in terms of tangential and cross-components of the
shear.  The negative sign is consistent with the spherical harmonic
definition~\eqref{eq:gamma-kappa}.
}
For validation, we show shear signals computed from the full-sky weak lensing
maps, so that we do not have to account for the effect of shot noise from the
galaxy positions, which is difficult to model theoretically at our intended
level of accuracy.  Another difficulty is the reduced shear approximation
\citep{2010A&A...523A..28K, 2020A&A...636A..95D}: galaxies trace the reduced
shear~\eqref{eq:g}, whereas theory codes in general only compute the angular
power spectrum of the convergence or shear field.  The difference between the
two cases is readily seen in our simulations, as shown in
Figure~\ref{fig:galaxies_err}.  For an accurate evaluation of our results, we
must therefore compare the shear, and not the reduced shear, with the
theoretical values computed by \emph{CAMB}.  Overall, we find very good
agreement at the sub-per cent level, in line with our expectations.

\section{Discussion \& Conclusions}
\label{sec:discussion}

We have introduced \emph{GLASS}, the Generator for Large Scale Structure, which
is a public code for creating simulations of wide-field galaxy surveys, with a
particular focus on weak gravitational lensing.  Our simulated light cones are
built as a series of nested matter shells around the observer, iteratively
sampled from a given statistical distribution.  If the matter field can be
approximated as uncorrelated beyond a certain length scale, which is a fair
approximation, our simulations can be carried out with constant memory use.
This allows us in principle to simulate any number of matter shells, and
therefore to achieve a much higher resolution than currently possible in both
the radial and angular dimensions.  As a result, our method readily achieves
per cent-level accuracy for clustering and weak lensing two-point statistics
for angular mode numbers~$l \gtrsim 3\,000$ and redshifts~$z \gtrsim 2$, which
are typical for Stage 4 photometric galaxy surveys.

A key part in that is a novel way to realise transformed Gaussian random fields,
such as e.g.\ lognormal fields, with angular power spectra of a given angular
range and practically arbitrary accuracy and precision.  Moreover, we developed
a scheme to compute the weak lensing convergence field iteratively, using a
multi-plane formalism usually employed in strong gravitational lensing.  The
accuracy of the weak lensing fields is essentially determined by the size of the
matter shells, and can therefore be controlled as necessary for a given
simulation.  The situation is similar for angular galaxy clustering, which is
more sensitive to the relative resolution of the matter shells compared to the
width of the galaxy redshift distribution.  Overall, the ability to increase the
radial resolution, and hence number of matter shells, without quadratically
increasing memory use, is therefore crucial.

\emph{GLASS} is fast: the high-precision matter, galaxy clustering, and lensing
simulations we present take around~30 minutes wall-clock time each on standard
8-core computing nodes, including analysis of the results.  Another benefit of
the iterative computation in shells is that results are available for processing
as soon as each new shell is computed.  Therefore, simulation and analysis
pipelines can be constructed in which no large amounts of data (e.g.\ catalogues
or maps) are ever written to disk.  This is particularly important since the
speed and resource efficiency of \emph{GLASS} can lead to input and output
becoming a limiting factor in such pipelines.

Our approach of a hybrid mix of statistical and physical models allows for
simulations in which each individual step is understandable, analysable, and
extensible, providing the simulator with control over the trade-off between
accuracy and speed/resource consumption.  The \emph{GLASS} design is completely
modular, and without a ``default mode'' of operation; all models we present in
this work, including the most basic ones for matter and lensing, are readily
replaced or expanded.  This makes \emph{GLASS} a well-suited tool for
stress-testing and validating the processing and analysis pipelines of galaxy
surveys.

We demonstrated that the \emph{GLASS} simulator matches or exceeds the accuracy
of our current analytic models of the dark matter distribution
\citep[c.f.][]{2019MNRAS.484.5509E,2021MNRAS.502.1401M}. Hence, simulation-based
inference of two-point statistics employing \emph{GLASS} will be at least as
accurate as traditional analytic approaches, but offers a much more
straightforward route to addressing otherwise formidable analysis challenges,
such as non-Gaussian likelihoods, higher-order signal corrections, complex
galaxy sample selection, and spatially varying survey properties, to name just a
few. In forthcoming work we will extend the \emph{GLASS} approach to also
produce highly accurate higher-order statistics of the matter distribution to
enable their simultaneous inference.

\paragraph{Acknowledgements}

We would like to thank A.~Hall for his always very helpful comments and
insights, as well as the anonymous reviewer for their constructive comments
which improved this text.

NT, AL, and BJ are supported by UK Space Agency grants ST/W002574/1 and
ST/X00208X/1. BJ is also supported by STFC Consolidated Grant ST/V000780/1. MvWK
acknowledges  STFC for support in the form of a PhD Studentship.

We gratefully acknowledge use of the following software packages: \emph{NumPy}
\citep{2020Natur.585..357H}, \emph{HEALPix} \citep{2005ApJ...622..759G},
\emph{healpy} \citep{2019JOSS....4.1298Z}, and \emph{Matplotlib}
\citep{2007CSE.....9...90H}.

\paragraph{Data availability}

All data and software used in this article is publicly available.  \emph{GLASS}
is open source software and its repository and documentation can be found
online.  The scripts to generate the simulations and plots presented here can be
found in a separate software repository.  All Python packages mentioned in the
text can be obtained from the Python Package Index.

\bibliographystyle{mnras}
\bibliography{main}

\setlength{\twocolwidth}{\columnwidth}

\appendix

\section{Discrete Legendre Transform}
\label{sec:app-a}

The transformation~\eqref{eq:cltoct} and~\eqref{eq:cttocl} between angular power
spectra and angular correlation functions is essentially the Legendre expansion
of the function~$C(\theta)$.  To compute it for inputs of finite length, we use
a recursive algorithm, based on the starting point for the algorithms of
\citet{1991SIAM...12..158A}.  The main idea is as follows.  If a function~$f$
has a finite Legendre expansion of the form
\begin{equation}
\label{eq:dlt-leg}
    f(\theta)
    = \sum_{l=0}^{n-1} a_l \, P_l(\cos\theta) \;,
\end{equation}
then it also has a finite Fourier cosine expansion of the form
\begin{equation}
\label{eq:dlt-cos}
    f(\theta)
    = \sum_{k=0}^{n-1} b_k \cos(k\theta) \;,
\end{equation}
since~$P_l(\cos\theta)$ is a polynomial of degree~$l$ in~$\cos\theta$.  The
coefficient vectors~$\vec{a} = \{a_0, \ldots, a_{n-1}\}$ and~$\vec{b} = \{b_0,
\ldots, b_{n-1}\}$ are related as $\vec{b} = \mat{M} \vec{a}$, where the
matrix~$\mat{M}$ has entries
\begin{equation}
\label{eq:dlt-m}
    M_{ij} = \begin{cases}
        \frac{1}{\pi} \,
            \frac{\Gamma(\frac{j+1}{2})^2}%
                 {\Gamma(\frac{j+2}{2})^2}
        & \text{if $0 = i \le j < n$ and $j$ even,}
        \\[10pt]
        \frac{2}{\pi} \,
            \frac{\Gamma(\frac{j-i+1}{2}) \, \Gamma(\frac{j+i+1}{2})}%
                 {\Gamma(\frac{j-i+2}{2}) \, \Gamma(\frac{j+i+2}{2})}
        & \text{if $0 < i \le j < n$ and $i+j$ even,}
        \\[10pt]
        0
        & \text{otherwise,}
    \end{cases}
\end{equation}
and~$\Gamma$ is the gamma function.  Conversely, if~$f$ has a finite Fourier
cosine expansion~\eqref{eq:dlt-cos}, then it also has a finite Legendre
expansion~\eqref{eq:dlt-leg}, and the coefficient vectors are related
as~$\vec{a} = \mat{L} \vec{b}$, where the matrix~$\mat{L}$ has entries
\begin{equation}
\label{eq:dlt-l}
    L_{ij} = \begin{cases}
        1
        & \text{if $i = j = 0$,}
        \\[10pt]
        \frac{\sqrt{\pi}}{2} \,
            \frac{\Gamma(\frac{2i+2}{2})}{\Gamma(\frac{2i+1}{2})}
        & \text{if $0 < i = j < n$,}
        \\[10pt]
        \frac{-j \, (i + \frac{1}{2})}{(j + i + 1) \, (j - i)} \,
            \frac{\Gamma(\frac{j-i-1}{2}) \, \Gamma(\frac{j+i}{2})}%
                 {\Gamma(\frac{j-i}{2}) \, \Gamma(\frac{j+i+1}{2})}
        & \text{if $0 \le i < j < n$ and $i + j$ even,}
        \\[10pt]
        0
        & \text{otherwise.}
    \end{cases}
\end{equation}
Since the transformation between coefficient vector~$\vec{b}$ and function
values~$f(\theta)$ can be done efficiently with a Discrete Cosine Transform
(DCT), the Discrete Legendre Transform (DLT) reduces to a DCT and matrix
multiplication $\vec{a} = \mat{L} \vec{b}$; and the inverse DLT reduces to a
matrix multiplication $\vec{b} = \mat{M} \vec{a}$ and inverse DCT.

With fast algorithms for the DCT widely available, our task reduces to computing
the matrix products with~$\mat{L}$ and~$\mat{M}$.   Although the main result of
\citet{1991SIAM...12..158A} was an efficient algorithm for this purpose, here we
use a simple recurrence to compute~$\mat{M}\vec{a}$ or~$\mat{L}\vec{b}$ without
explicitly constructing a large matrix.  The recursive computation does not have
the same algorithmic complexity as the method proposed by
\citet{1991SIAM...12..158A}, but is nevertheless very fast due to its
simplicity.

The entries~\eqref{eq:dlt-m} of~$\mat{M}$ can alternatively be specified by the
first two diagonal elements $M_{00} = M_{11} = 1$, from which all subsequent
diagonal elements~$M_{ii}$, $i > 1$, can be computed as
\begin{equation}
    M_{ii}
    = \Bigl(1 - \frac{1}{2i}\Bigr) \, M_{i-1,i-1} \;.
\end{equation}
The lower triangle of the matrix is identically zero.  Above the diagonal, the
values~$M_{ij}$ for $j > i$ can be computed as
\begin{equation}
    M_{ij}
    = \Bigl(1 - \frac{1}{j-i}\Bigr) \,
        \Bigl(1 - \frac{1}{j+i}\Bigr) \, M_{i,j-2} \;.
\end{equation}
The first off-diagonal and every other subsequent entry vanishes ($i + j$ odd).

The entries~\eqref{eq:dlt-l} of~$\mat{L}$ can be computed similarly by starting
the diagonal with $L_{00} = L_{11} = 1$ and continuing as
\begin{equation}
    L_{ii}
    = \frac{1}{1 - \frac{1}{2i}} \, L_{i-1,i-1} \;.
\end{equation}
The lower triangle is again zero, and the values~$L_{ij}$ for $j > i$ can be
computed as
\begin{equation}
    L_{ij}
    = \Bigl(1 + \frac{2}{j-2}\Bigr) \, \Bigl(1 - \frac{3}{j-i}\Bigr) \,
        \Bigl(1 - \frac{3}{j+i+1}\Bigr) \, L_{i,j-2} \;.
\end{equation}
As before, the first off-diagonal and every other subsequent entry vanishes ($i
+ j$ odd).

\section{Iterative multivariate normal random sampling}
\label{sec:app-b}

Let~$\vec{x}_{n+1} = \{x_1, \ldots, x_{n+1}\}$ be a multivariate normal random
vector of length~$n+1$ with mean~$\vec{\mu}_{n+1} = \{\mu_1, \ldots,
\mu_{n+1}\}$ and covariance matrix~$\mat{\Sigma}_{n+1}$.  The vector~$\vec{x}_n$
of the first~$n$ variates is then a multivariate normal random vector with
mean~$\vec{\mu}_n$ and covariance matrix~$\mat{\Sigma}_n$, which is the leading
$n \times n$ submatrix of~$\mat{\Sigma}_{n+1}$.  Given a sample~$\vec{x}_n$ from
its marginal distribution, what is the conditional distribution of~$x_{n+1}$?

Of course, this is a classical problem with a well-known solution:  Writing the
covariance matrix~$\mat{\Sigma}_{n+1}$ in block matrix form as
\begin{equation}
\label{eq:iter-cov}
    \mat{\Sigma}_{n+1}
    = \begin{bmatrix}
        \mat{\Sigma}_n & \vec{c}_n \\
        \vec{c}_n^\tp & \sigma^2_{n+1}
    \end{bmatrix} \;,
\end{equation}
the conditional distribution of~$x_{n+1}$ is normal with mean
\begin{equation}
\label{eq:iter-mu}
    \tilde{\mu}_{n+1}
     = \mu_{n+1}
     + \vec{c}_n^\tp \, \mat{\Sigma}_n^{-1} \, (\vec{x}_n - \vec{\mu}_n)
\end{equation}
and variance
\begin{equation}
\label{eq:iter-var}
    \tilde{\sigma}^2_{n+1}
    = \sigma^2_{n+1}
    - \vec{c}_n^\tp \, \mat{\Sigma}_n^{-1} \, \vec{c}_n \;.
\end{equation}
We can hence sample from a multivariate normal distribution by sampling each
individual normal variate in turn, using the conditional mean~\eqref{eq:iter-mu}
and variance~\eqref{eq:iter-var}.

In fact, the entire sampling process can be performed iteratively.
Let~$\mat{A}_n$ be the Cholesky decomposition of~$\mat{\Sigma}_n^{-1}$ such
that~$\mat{\Sigma}_n^{-1} = \mat{A}_n^\tp \, \mat{A}_n$.  We can write the
conditional mean~\eqref{eq:iter-mu} and variance~\eqref{eq:iter-var} as
\begin{gather}
\label{eq:iter-mu:2}
    \tilde{\mu}_{n+1}
    = \mu_{n+1} + \vec{a}_n^\tp \, \vec{y}_n \;,
\\
\label{eq:iter-var:2}
    \tilde{\sigma}^2_{n+1}
    = \sigma^2_{n+1} - \vec{a}_n^\tp \, \vec{a}_n \;,
\end{gather}
where we have introduced vectors~$\vec{a}_n = \mat{A}_n \, \vec{c}_n$
and~$\vec{y}_n = \mat{A}_n \, (\vec{x}_n - \vec{\mu}_n)$.
Since~$\mat{A}_n^\tp \, \mat{A}_n = \mat{\Sigma}_n^{-1}$, the vector~$\vec{y}_n$
contains the~$n$ standard normal random variates of the whitened given sample.
It is updated by appending each new standard normal variate~$y_{n+1} =
(x_{n+1} - \tilde{\mu}_{n+1})/\tilde{\sigma}_{n+1}$ as it is drawn,
\begin{equation}
    \vec{y}_{n+1}
    = \begin{pmatrix}
        \vec{y}_{n} \\
        y_{n+1}
    \end{pmatrix}  \;.
\end{equation}
To obtain the conditional mean and variance, we therefore only have to compute
the vector~$\vec{a}_n$ explicitly.  To do so, we require~$\mat{A}_n$.

Writing the matrix $\mat{\Sigma}_n$ in block matrix form~\eqref{eq:iter-cov},
the inverse~$\mat{\Sigma}_n^{-1}$ can be computed using block matrix inversion,
and factorised as
\begin{equation}
\label{eq:iter-cov-upd}
    \mat{\Sigma}_n^{-1}
    = \begin{pmatrix}
        \mat{I} & -\mat{\Sigma}_{n-1}^{-1} \, \vec{c}_{n-1} \\
        \mat{0} & \mat{I}
    \end{pmatrix} \, \begin{pmatrix}
        \mat{\Sigma}_{n-1}^{-1} & \mat{0} \\
        \mat{0} & \tilde{\sigma}^{-2}_n
    \end{pmatrix} \, \begin{pmatrix}
        \mat{I} & \mat{0} \\
        -\vec{c}_{n-1}^\tp \, \mat{\Sigma}_{n-1}^{-1} & \mat{I}
    \end{pmatrix} \;,
\end{equation}
where~$\mat{I}$ and~$\mat{0}$ are the identity and zero matrix, respectively, of
the appropriate shape.  Looking at~\eqref{eq:iter-cov-upd}, the updating rule
for~$\mat{A}_n$ is readily obtained,
\begin{equation}
\label{eq:iter-a}
    \mat{A}_n
    = \begin{pmatrix}
        \mat{A}_{n-1} &
        \mat{0} \\
        -\tilde{\sigma}_n^{-1} \, \vec{a}_{n-1}^\tp \, \mat{A}_{n-1} &
        \tilde{\sigma}_n^{-1}
    \end{pmatrix} \;.
\end{equation}
It is clear that~$\mat{A}_n$ grows by one row and one column in each iteration,
and is a lower triangular matrix for all~$n$.

To sample~$x_{n+1}$ iteratively, we hence need to store the~$n$ standard normal
variates~$\vec{y}_n$ and the~$n \times n$ matrix~$\mat{A}_n$.  However, the
storage requirements reduce from~$n$ to~$k$ if each new random
variable~$x_{n+1}$ only correlates with the last~$k$ random variates $x_{n-k+1},
\ldots, x_n$.  The covariance matrix~$\mat{\Sigma}_{n+1}$ is then a banded
matrix, so that the vector~$\vec{c}_n$ has $n - k$ leading zeros.  In block
matrix form,
\begin{equation}
    \vec{c}_n
    = \begin{pmatrix}
        \vec{0} \\
        \tilde{\vec{c}}_n
    \end{pmatrix} \;,
\end{equation}
where~$\tilde{\vec{c}}_n$ is the reduced vector of~$k$ non-zero correlations.
The lower triangular matrix~$\mat{A}_n$ can be written in block matrix form as
\begin{equation}
    \mat{A}_n
    = \begin{pmatrix}
        \mat{U}_n & \mat{0} \\
        \mat{V}_n & \tilde{\mat{A}}_n
    \end{pmatrix} \;,
\end{equation}
where~$\tilde{\mat{A}}_n$ is the lower $k \times k$ block of~$\mat{A}_n$,
with~$\mat{U}_n$, $\mat{V}_n$ the remaining blocks.  It follows that the
vector~$\vec{a}_n = \mat{A}_n \, \vec{c}_n$ will also have $n - k$ leading
zeros, since
\begin{equation}
    \vec{a}_n
    = \begin{pmatrix}
        \mat{U}_n & \mat{0} \\
        \mat{V}_n & \tilde{\mat{A}}_n
    \end{pmatrix} \, \begin{pmatrix}
        \vec{0} \\
        \tilde{\vec{c}}_n
    \end{pmatrix}
    = \begin{pmatrix}
        \vec{0} \\
        \tilde{\mat{A}}_n \, \tilde{\vec{c}}_n
    \end{pmatrix}
    = \begin{pmatrix}
        \vec{0} \\
        \tilde{\vec{a}}_n
    \end{pmatrix} \;,
\end{equation}
where~$\tilde{\vec{a}}_n = \tilde{\mat{A}}_{n} \, \tilde{\vec{c}}_n$ is the
non-zero part of~$\vec{a}_n$ of length~$k$.  The conditional
mean~\eqref{eq:iter-mu:2} and variance~\eqref{eq:iter-var:2} can hence be
written
\begin{gather}
\label{eq:iter-mu:3}
    \tilde{\mu}_{n+1}
    = \mu_{n+1} + \tilde{\vec{a}}_n^\tp \, \tilde{\vec{y}}_n \;,
\\
\label{eq:iter-var:3}
    \tilde{\sigma}^2_{n+1}
    = \sigma^2_{n+1} - \tilde{\vec{a}}_n^\tp \, \tilde{\vec{a}}_n \;,
\end{gather}
where~$\tilde{\vec{y}}_n$ is the reduced vector of the last~$k$ standard normal
variates.  Writing the update~\eqref{eq:iter-a} of the matrix~$\mat{A}_n$ in
terms of reduced quantities,
\begin{equation}
    \mat{A}_n
    = \begin{pmatrix}
        \mat{U}_{n-1} &
        \mat{0} &
        \mat{0} \\
        \mat{V}_{n-1} &
        \tilde{\mat{A}}_{n-1} &
        \mat{0} \\
        -\tilde{\sigma}_n^{-1} \, \tilde{\vec{a}}_{n-1}^\tp \, \mat{V}_{n-1} &
        -\tilde{\sigma}_n^{-1} \, \tilde{\vec{a}}_{n-1}^\tp \,
                                                        \tilde{\mat{A}}_{n-1} &
        \tilde{\sigma}_n^{-1}
    \end{pmatrix} \;,
\end{equation}
it is clear that only the lower $k \times k$ submatrix~$\tilde{\mat{A}}_n$ needs
to be computed and stored.  Its update can be written as
\begin{equation}
    \begin{pmatrix}
        \;\cdot\; & \cdots \\
        \vdots & \tilde{\mat{A}}_n
    \end{pmatrix}
    = \begin{pmatrix}
        \tilde{\mat{A}}_{n-1} &
        \mat{0} \\
        -\tilde{\sigma}_n^{-1} \, \tilde{\vec{a}}_{n-1}^\tp \,
                                                        \tilde{\mat{A}}_{n-1} &
        \tilde{\sigma}_n^{-1}
    \end{pmatrix} \;,
\end{equation}
where the dots on the left-hand side indicate the first row and column, which
can be discarded.  Hence, for iterative multivariate normal random sampling
of~$k$ correlated variates, we only need to store $k \, (k+3)/2$ numbers: the
$k$ vector~$\tilde{\vec{y}}_n$, and the $k \times k$ lower triangular
matrix~$\tilde{\mat{A}}_n$.

\section{Normal distributions for galaxy ellipticity}
\label{sec:app-c}

The galaxy ellipticity components~$\epsilon_1, \epsilon_2$ are commonly sampled
as bivariate normal random variates with zero mean, variance~$\sigma_\epsilon^2$
in each component, and no correlation,
\begin{equation}
\label{eq:eps-normal}
    \epsilon_1, \epsilon_2 \sim \mathcal{N}(0, \sigma_\epsilon^2) \;.
\end{equation}
We call this the ``extrinsic'' normal distribution of ellipticities,
because~$\epsilon$ is treated here as a point in the Euclidean plane, with no
regard for e.g.\ the requirement that~$|\epsilon| \le 1$.  To obtain valid
ellipticities, we must reject samples with~$|\epsilon| > 1$ explicitly after the
fact.

It is also possible to define a normal distribution that produces valid
ellipticities naturally.  To do so, we first note that the space of
ellipticities is not the Euclidean plane but the hyperbolic plane; more
specifically, the ellipticity~$\epsilon$ denotes a point in the Poincar\'e disc
model of hyperbolic geometry \citep[see e.g.][]{1997hg...book.....C}.  While
this can be shown rigorously \citep[see e.g.\ Chapter 2.3
of][]{2009epde.book.....A}, here we will only point to the fact that the
transformation law~\eqref{eq:eps-wl} of weak lensing is precisely an isometric
translation of the hyperbolic plane which maps the origin to~$g$.

The hyperbolic plane is a Riemannian manifold, and we may therefore use the
method of \citet{2006JMIV..25...127P} to define an ``intrinsic'' normal
distribution for ellipticities:  It is the normal (in the statistical sense)
distribution in normal (in the differential-geometric sense) coordinates in the
mean of the distribution.  Since the mean of an isotropic ellipticity
distribution is necessarily zero, let~$\eta = \eta_1 + \I \, \eta_2$ be normal
coordinates in the origin of the hyperbolic plane.  Writing~$\eta$ in polar
representation,
\begin{equation}
    \eta = |\eta| \, \E^{2\I \phi} \;,
\end{equation}
the angle~$\phi$ is the same as that of its associated ellipticity~$\epsilon$
in~\eqref{eq:eps-isophot}, and the magnitude~$|\eta|$ is related to~$|\epsilon|$
as
\begin{equation}
\label{eq:eps-norm}
    |\eta|
    = 2 \atanh |\epsilon|
    = -2\ln q \;.
\end{equation}
The intrinsic normal distribution for ellipticities can therefore be sampled
straightforwardly by drawing uncorrelated normal random variates~$\eta_1,
\eta_2$ instead of~$\epsilon_1, \epsilon_2$,
\begin{equation}
    \eta_1, \eta_2 \sim \mathcal{N}(0, \sigma_\eta^2) \;,
\end{equation}
where~$\sigma_\eta^2$ is now the per-component variance in normal coordinates,
and inverting the relation~\eqref{eq:eps-norm} to convert the normal coordinates
back to the associated ellipticity~$\epsilon$.

\begin{figure*}%
\begin{minipage}[t]{\twocolwidth}%
\centering%
\includegraphics[scale=0.85]{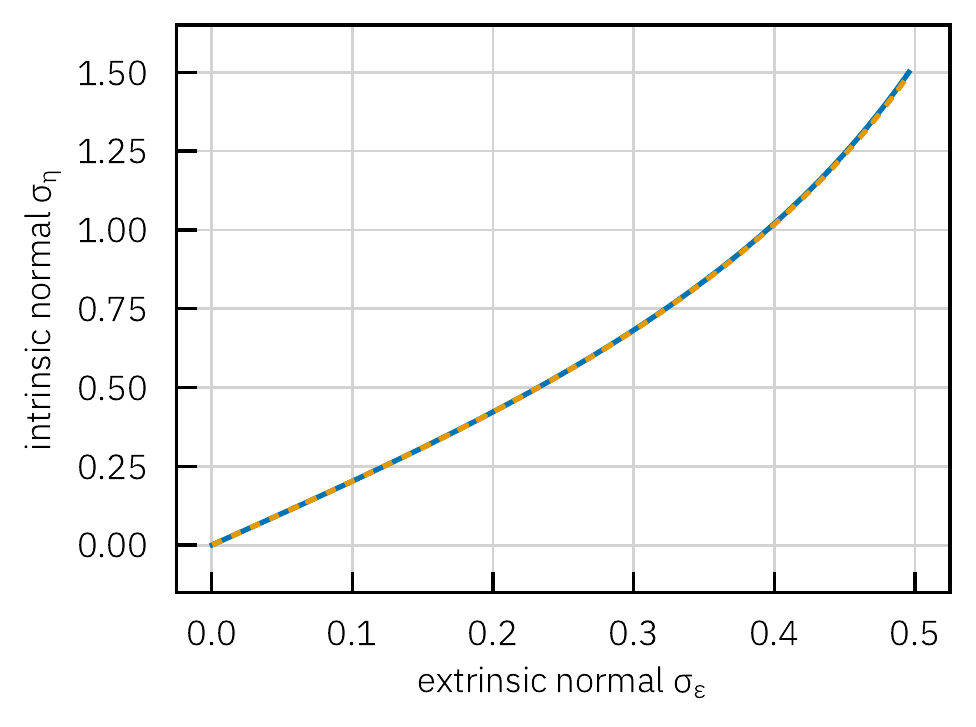}%
\caption{%
Relation between the standard deviations~$\sigma_\epsilon$ and~$\sigma_\eta$ of
the extrinsic and intrinsic normal distribution for ellipticity (\emph{blue})
and a fit by a rational function (\emph{dotted orange}).
}%
\label{fig:intnorm_var}%
\end{minipage}%
\hfill%
\begin{minipage}[t]{\twocolwidth}%
\centering%
\includegraphics[scale=0.85]{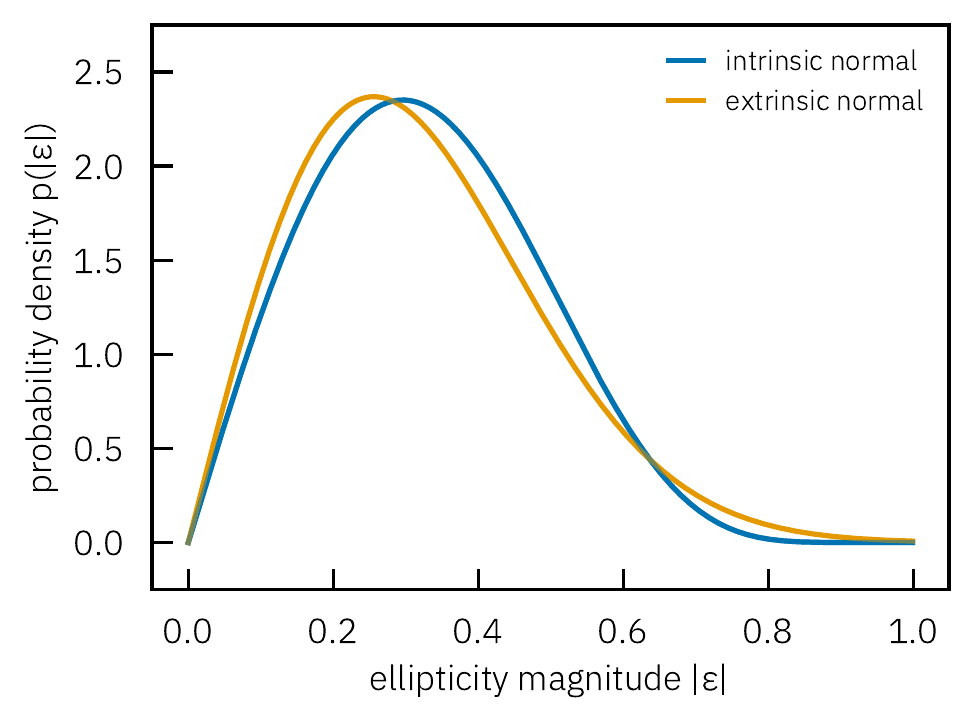}%
\caption{%
Comparison of intrinsic normal and extrinsic normal ellipticity distributions
with standard deviation~$\sigma_\epsilon = 0.256$ in each component.
}%
\label{fig:intnorm}%
\end{minipage}%
\end{figure*}

The change of variable~\eqref{eq:eps-norm} yields the probability distribution
function~$p$ of the ellipticity magnitude~$|\epsilon|$ under the intrinsic
normal distribution,
\begin{equation}
\label{eq:p-eps-intr}
    p(|\epsilon|)
    = \frac{4 \, \E^{-\frac{2 \atanh^2 |\epsilon|}{\sigma_\eta^2}}
                                                    \atanh |\epsilon|}
           {\sigma_\eta^2 \, (1 - |\epsilon|^2)} \;.
\end{equation}
In particular, the resulting standard deviation~$\sigma_\epsilon$ for a
given~$\sigma_\eta$ can be computed using~\eqref{eq:p-eps-intr}, which yields
the relation shown in Figure~\ref{fig:intnorm_var}.  A useful fit to the
curve is
\begin{equation}
    \sigma_\eta^2
    \approx \sigma_\epsilon^2 \, \frac{8 + 5 \, \sigma_\epsilon^2}
                                      {2 - 4 \, \sigma_\epsilon^2} \;,
\end{equation}
which is good enough to match the intrinsic normal galaxy ellipticity
distribution to data from a survey.  A comparison between the intrinsic normal
and extrinsic normal distributions with~$\sigma_\epsilon = 0.256$ is shown in
Figure~\ref{fig:intnorm}.  The most notable difference is the more realistic
suppression of high ellipticities with $|\epsilon| > 0.8$ in the intrinsic
normal case.

Besides, there is also a computational advantage: The extrinsic normal
distribution requires explicit rejection of ellipticities with~$|\epsilon| > 1$,
and hence repeated sampling, particularly when~$\sigma_\epsilon^2$ is large.
The intrinsic normal distribution accepts all samples for any value of the
variance~$\sigma_\epsilon^2$.  Given the slightly more realistic shape of the
distribution, as well as the improved sampling efficiency, we can generally
recommend the intrinsic normal distribution for galaxy ellipticities,
particularly when their numbers are large.

\end{document}